\documentclass[prd,aps,twocolumn,showpacs,superscriptaddress,nofootinbib]{revtex4-1}
\usepackage{graphicx}
\usepackage{epsfig}
\usepackage{amsmath}
\usepackage{hyperref}
\usepackage{multirow}
\usepackage{amssymb}
\usepackage{subfigure}
\usepackage{float}
\usepackage{graphics}
\usepackage{slashed}
\usepackage{booktabs}
\usepackage{color}
\usepackage[normalem]{ulem}

\def\be{\begin{equation}}
\def\ee{\end{equation}}
\newcommand{\bea}{\begin{equation} \begin{array}{c}}
\newcommand{\eea}{ \end{array} \end{equation}}

\def\as{\alpha_s}

\def\nno{\nonumber}

\def\ff{f\hspace{-0.4em}f}
\def\II{I\hspace{-0.7em}I}
\def\ptv{p_T^{\rm veto}}

\def\hlinew#1{%
  \noalign{\ifnum0=`}\fi\hrule \@height #1 \futurelet
   \reserved@a\@xhline}
\begin{document}

\title{Some recent theoretical progress in Higgs boson and top quark physics at hadron colliders}

\author{Chong Sheng Li}
\email{csli@pku.edu.cn}
\affiliation{School of Physics and State Key Laboratory of Nuclear Physics and Technology, Peking University, Beijing 100871, China}
\affiliation{Center for High Energy Physics, Peking University, Beijing 100871, China}
\author{Hai Tao Li}
\affiliation{School of Physics and State Key Laboratory of Nuclear Physics and Technology, Peking University, Beijing 100871, China}
\author{Ding Yu Shao}
\affiliation{School of Physics and State Key Laboratory of Nuclear Physics and Technology, Peking University, Beijing 100871, China}

\date{\today}

\begin{abstract}

The test of the Standard Model and search for New Physics signal are main aim of LHC experiment. With the increasing of the measurement accuracy at the LHC, it is a major task in future to exceed the current accuracy of the theoretical predictions for important processes,
in particular ones involving Higgs boson and top quark. In this review we briefly summarize some recent theoretical progress in Higgs boson and top quark physics,
especially the fixed-order and resummation predictions in QCD at both the Tevatron and the LHC.
\end{abstract}

\maketitle
%\tableofcontents

\section{Introduction}
Recently, a  Higgs boson with a mass around $125~{\rm GeV}$ has been discovered by the ATLAS~\cite{Aad:2012tfa} and CMS~\cite{Chatrchyan:2012ufa} collaborations at the LHC. In the future, it is possible that the LHC can tell whether this particle is the Standard Model~(SM) Higgs boson or one of many Higgs bosons in new physics~(NP) model.

In the SM, the Higgs boson is responsible for the origin of Electro-Weak~(EW) symmetry breaking and the generation of elementary particle masses. The future experimental task at the LHC is to examine the Higgs mechanism and test the properties and couplings of Higgs boson. Therefore, in order to compare with more precise experimental results, it is important to perform accurate theoretical predictions for the Higgs process at the LHC.

Besides discovering Higgs boson, another important task at the LHC is the measurement of the top quark properties. In fact, the LHC has produced over a millon and around ten million top quark events at a center of mass energy of 7~TeV and 8~TeV, respectively, which leads to precise measurements of observables relevant to top quark. Thus, the accurate theoretical predictions are necessary in order to test the SM and search for NP.

In general, QCD controls the theoretical predictions for the production of any particle in both the SM and NP at hadron colliders. And the QCD high order corrections play a key role for the accurate theoretical predictions. These QCD corrections may come from  virtual corrections and extra hard parton emissions which involve complicated multi-loop and  multi-leg calculations, respectively. Besides, significant contributions can also come from the logarithmic terms by emitting the soft and collinear gluons, which can be resummed to all order in $\as$. A lot of efforts on QCD high order calculations have been made for over twenty years, and the theoretical predictions become more and more precise. In this review some recent theoretical progress in the Higgs and top quark physics are summarized below. 

\section{Recent progress in Higgs boson physics}
Recently, the important experimental results are the measurements of signal strength parameters of the Higgs boson at the LHC, reported by ATLAS and CMS collaborations. The rates of Higgs boson production and decay are parameterized using signal strength parameters $\mu$, which is defined as
\begin{eqnarray}\label{eq1}
 \mu = \frac{\sigma \times {\rm Br}}{ \left( \sigma \times {\rm Br} \right)_{\rm SM} }.
\end{eqnarray}

\begin{figure}[h]
  \centering
  \includegraphics[scale=0.3]{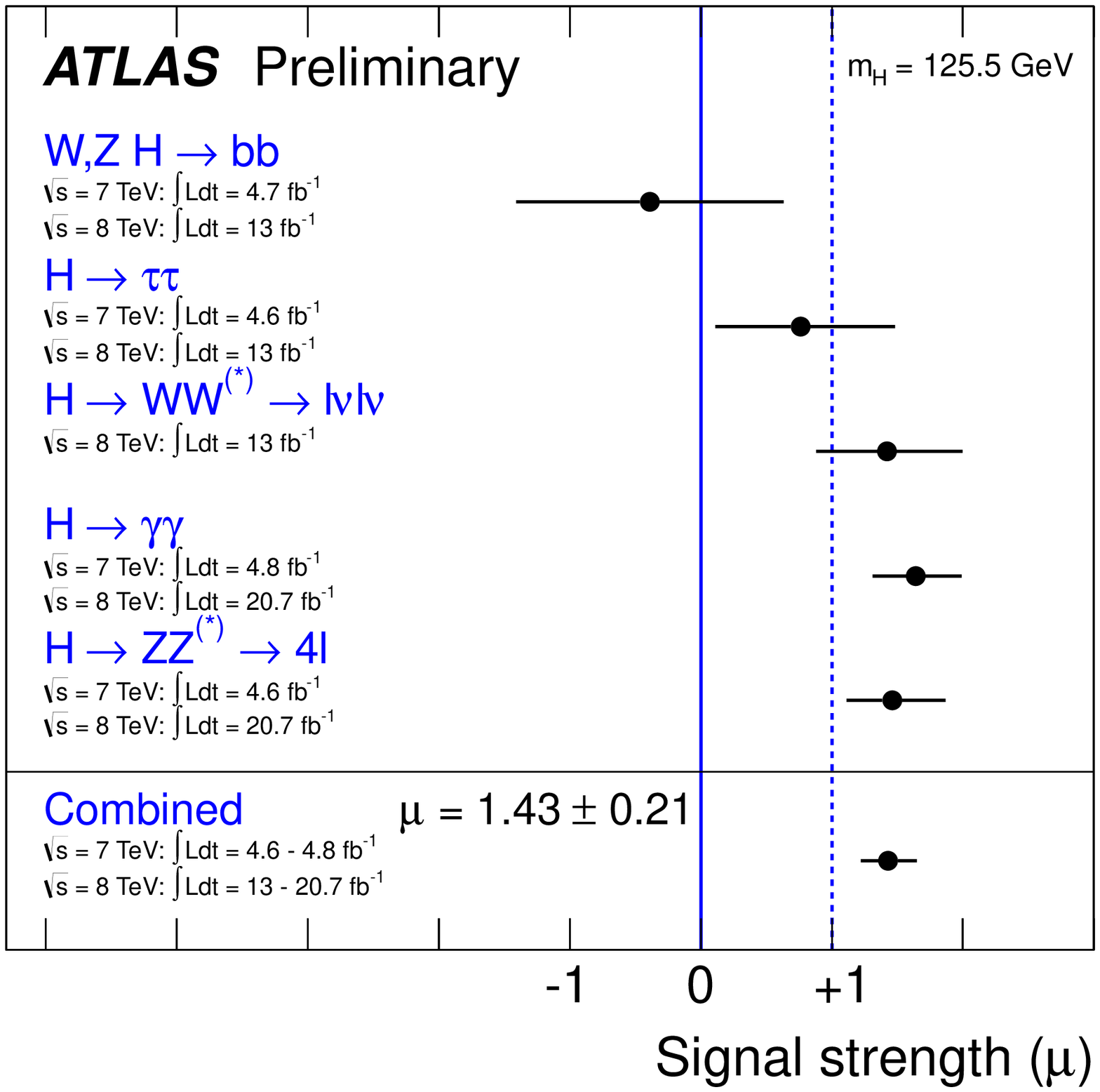}
  \quad
  \vspace{-2ex}
  \includegraphics[scale=0.3]{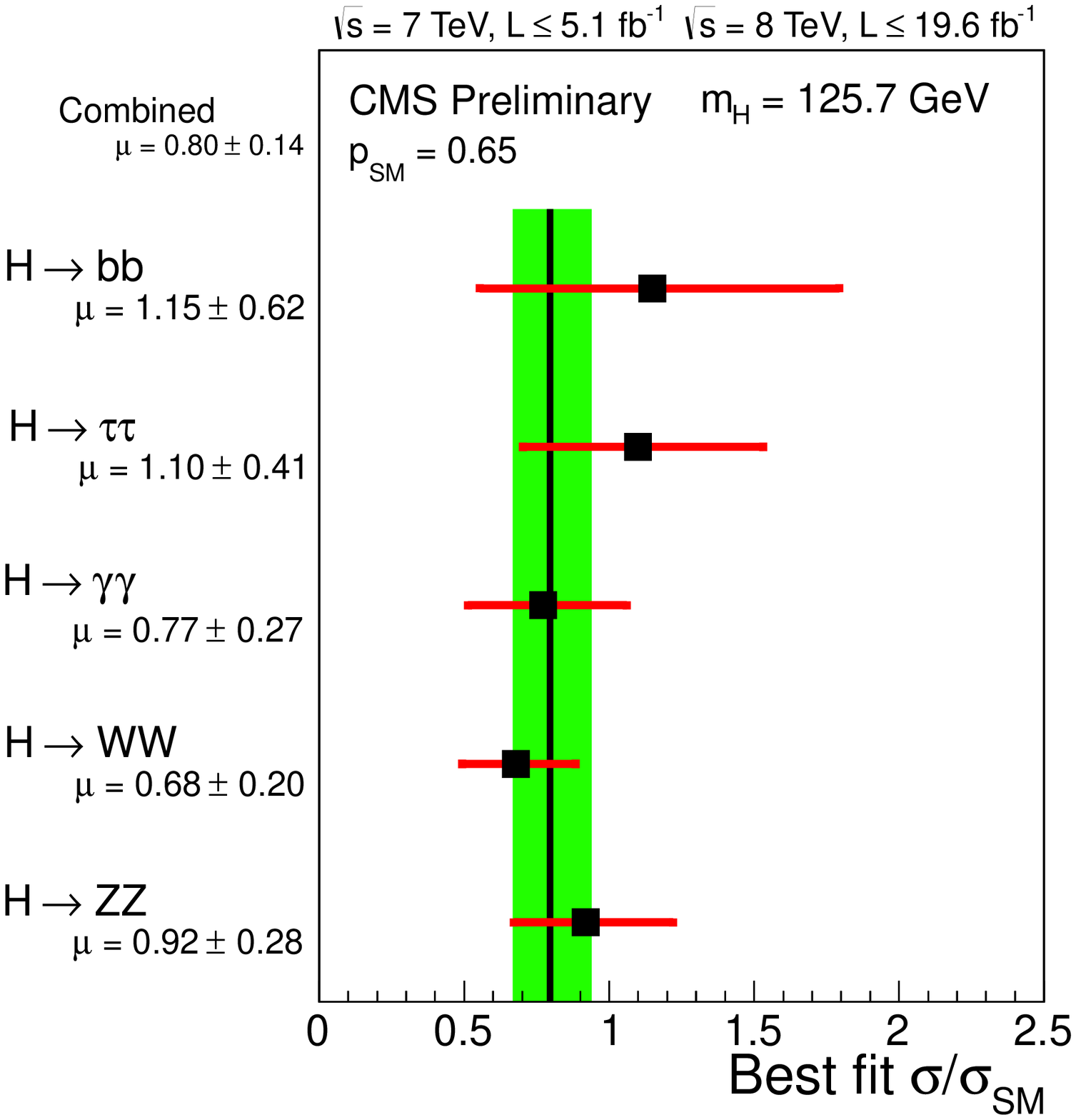}
  \vspace{-2ex}
  \caption{\label{fig:signal_strength} (Color online) Measurements of the signal strength parameter $\mu$ for the individual channels and their combinations~\cite{ATLAS-CONF-2013-034,CMS-PAS-HIG-13-005} }
\end{figure}

In Fig.~\ref{fig:signal_strength}, the signal strength for the various decay channels is shown by the ATLAS and CMS collaborations. For the ATLAS detector the combined signal strength is $\mu = 1.30 \pm 0.13({\rm stat.}) \pm 0.14({\rm syst.})$~\cite{ATLAS-CONF-2013-034}. For the CMS detector the combined signal strength is $\mu = 0.80 \pm 0.14$~\cite{CMS-PAS-HIG-13-005}. Obviously, as shown in Eq.~(\ref{eq1}), the parameter $\mu$ strongly depends on the accurate theoretical predictions at the LHC, especially the QCD predictions.

\subsection{Higgs boson production}

\begin{figure}[h]
  \centering
  \includegraphics[scale=0.4]{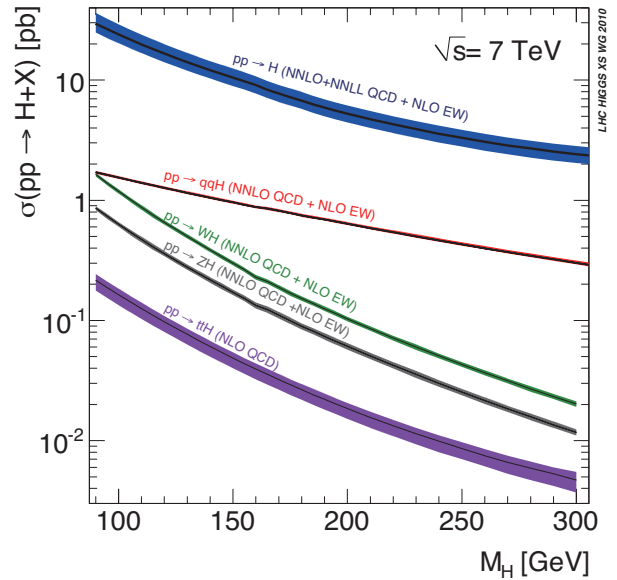} \hspace{0.5em}
  \vspace{-2ex}
  \caption{\label{fig:Higgs_XS_7TeV} (Color online) SM Higgs boson production cross sections at the LHC with $\sqrt{S}=7$ TeV~\cite{Dittmaier:2011ti,Dittmaier:2012vm,Heinemeyer:2013tqa}}
\end{figure}

At the LHC the SM Higgs boson is produced through four different channels:
\begin{itemize}
  \item Gluon gluon fusion channel: $gg \rightarrow hX$;
  \item Vector Boson Fusion~(VBF) channel: $qq' \rightarrow hjjX$;
  \item Higgs boson strahlung channel:  $q\bar{q} \rightarrow hVX$;
  \item Higgs boson and top quark pair associated production channel : $q\bar{q}(gg) \rightarrow ht\bar{t}X$.
\end{itemize}
Until now, the precision predictions for above cross sections at the LHC with $\sqrt{S}=7$ TeV are shown in Refs. \cite{Dittmaier:2011ti,Dittmaier:2012vm,Heinemeyer:2013tqa} (see Fig.~\ref{fig:Higgs_XS_7TeV}).

\subsubsection{Gluon gluon fusion channel}

\begin{figure}[h]
  \centering
  \includegraphics[scale=0.5]{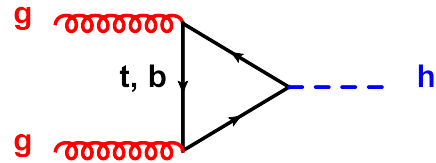} \hspace{0.5em}
  \vspace{-2ex}
  \caption{\label{fig:gg2H} (Color online) Feynman diagram of $gg\rightarrow h$ at the LO}
\end{figure}

Gluon gluon fusion induced by top and bottom quark loops is the dominating channel of Higgs boson production at the LHC (Fig.~\ref{fig:gg2H}), where the main contributions come from top quark loop due to the large Yukawa coupling $y_t \sim 1$. The QCD Next-to-Leading Order (NLO) corrections to this process have been investigated in both cases of the infinite~\cite{Dawson:1990zj,Djouadi:1991tka} and finite~\cite{Graudenz:1992pv,Spira:1995rr} top quark mass limits, and can enhance the total cross section by about 80~\% for a $125$ GeV Higgs boson at the LHC with $\sqrt{S}=7$ TeV. In the infinite top quark mass limit, the QCD Next-to-Next-to-Leading Order (NNLO) corrections to the total and differential cross section have been calculated~\cite{Harlander:2002wh,Anastasiou:2002yz,Ravindran:2003um,Anastasiou:2005qj,Catani:2007vq,Anastasiou:2007mz}, and increase the NLO results by about $25~\%$. Recently, in Ref.~\cite{Ball:2013bra}, approximate results for the total cross section at QCD Next-to-Next-to-Next-to-
Leading Order (N$^3$LO) in infinite top mass are also
calculated, and
the results show that the approximate N$^3$LO result amounts to a correction of $17~\%$ to the QCD NNLO cross section for a 125~GeV Higgs at the LHC with 8~TeV. Furthermore, the threshold soft gluon effects have been resummed up to Next-to-Next-to-Leading Logarithm (NNLL)~\cite{Catani:2003zt}, leading to an increase of cross section by about $7~\%$ at the LHC. And the Next-to-Next-to-Next-to-Leading Logarithm  resummation has also been studied~\cite{Moch:2005ky,Idilbi:2005ni,Ahrens:2008nc,Ahrens:2010rs}. Moreover, in Soft Collinear Effective Theory (SCET) formalism, the $\pi^2$ enhancement contributions, originating from the powers of logarithmic terms of $\ln[(-Q^2-{\rm i}\epsilon)/\mu_f^2]$, have been studied and resummed to all order~\cite{Ahrens:2008qu,Ahrens:2008nc,Ahrens:2010rs}, which help to explain the poor convergence behavior of fixed-order calculations.

\begin{figure}[h]
  \centering
  \includegraphics[scale=0.4]{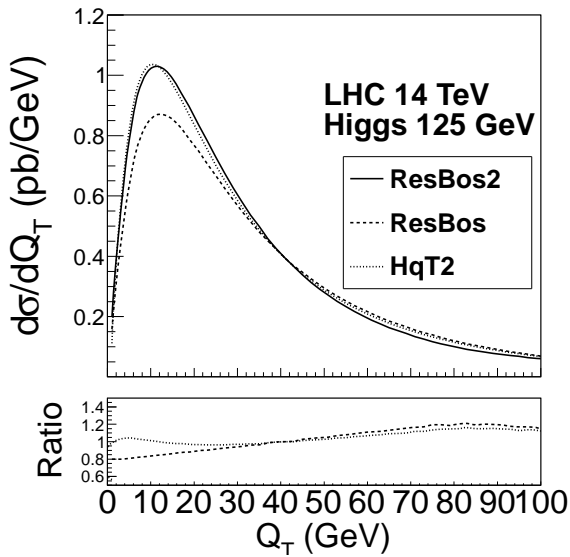} \hspace{0.5em}
  \vspace{-2ex}
  \caption{\label{fig:COMPARE_LHC14}   Different theoretical predictions on the transverse momentum distributions for the Higgs boson production at the 14 TeV LHC. In the bottom of each plot, the ratios to ResBos2 predictions are also shown~\cite{Wang:2012xs}}
\end{figure}

Besides the total cross section, the transverse momentum ($p_{\rm T}$) spectrum of Higgs boson can help to improve the statistical significance, especially in small $p_{\rm T}$ region. Up to $\mathcal{O}({\alpha_{\rm s}^4})$, in the large $p_{\rm T}$ region ($p_{\rm T} \gg m_h$) the $p_{\rm T}$ spectrum has been investigated in Refs.~\cite{Kauffman:1996ix,deFlorian:1999zd,Ravindran:2002dc,
Glosser:2002gm,Anastasiou:2005qj,Catani:2007vq,Grazzini:2008tf}. Moreover, contributions at $\mathcal{O}({\alpha_{\rm s}^5})$ to the production of Higgs boson associated with two jets have also been accomplished in Ref.~\cite{Campbell:2006xx}. However, in the small $p_{\rm T}$ region ($p_{\rm T} \ll m_h$), the convergence of the fixed-order expansion is spoiled by the large logarithmic terms $\ln(m_h^2/p_{\rm T}^2)$. In order to obtain reliable predictions, these logarithmic terms have to be resummed to all orders, which was done in Refs.~\cite{Yuan:1991we,Balazs:2000wv,Balazs:2000sz,Berger:2002ut,
Kulesza:2003wn,Bozzi:2003jy,Bozzi:2005wk,Bozzi:2007pn,Cao:2007du,
Cao:2009md,deFlorian:2011xf,Becher:2012qa,Becher:2012yn}. Recently, in the Collins-Soper-Sterman (CSS) framework~\cite{Collins:1981uk,Collins:1981va,Collins:1984kg}, the improved resummation calculations for the Higgs boson production via gluon gluon fusion by including the NNLO Wilson coefficient functions and G-functions have been completed. And the corresponding results are included in the ResBos program~\cite{Wang:2012xs}. The resummation formula can be written as
\begin{align}
 \label{eq:resum_formalism}
 & \frac{ {\rm d}\sigma(gg\to HX ) }{ {\rm d}Q^2 {\rm d}Q_{\rm T}^2 {\rm d}y }
  = \kappa \sigma_0 \frac{Q^2}{S} \frac{Q^2 \mathnormal{\Gamma}_H/m_H}{(Q^2-m_H^2)^2+(Q^2\mathnormal{\Gamma}_H/m_H)^2}
\nonumber\\
 & \times \Biggl\{ \frac{1}{(2\pi)^2} \int {\rm d}^2b {\rm e}^{{\rm i}Q_{\rm T}\cdot b}
\widetilde{W}_{gg}(b_*,Q,x_1,x_2,C_{1,2,3})
\nonumber\\
&\times\widetilde{W}_{gg}^{\rm NP}(b,Q,x_1,x_2)+Y(Q_{\rm T},Q,x_1,x_2,C_4)\Biggr\},
\end{align}
where the updated NNLO Wilson coefficients are included in $\widetilde{W}_{gg}$, which dominates at small $Q_{\rm T}$,
 and behaves as $Q_{T}^{-2}$ times a series of $\ln^n{(Q^2/Q_{\rm T}^2)}$. The function $\widetilde{W}_{gg}^{\rm NP}$
 describes the non-perturbative part, and the term containing $Y$ incorporates the remainder of the cross section
 which is not singular as $Q_{\rm T}\rightarrow 0$. The results show that including NNLO Wilson coefficient functions
 increases the total cross section predictions of ResBos for a 125 GeV Higgs Boson production by about 8~\% and 6\% at
  the Tevatron and the LHC, respectively. The different theoretical predictions on the transverse momentum distributions
  for the Higgs boson production at the LHC wiht 14~TeV are shown in Fig~\ref{fig:COMPARE_LHC14}.
%The improved ResBos (ResBos2) program includes the Wilson coefficient function up to NNLO and the Sudakov exponent with  NNLL accuracy, and is matched to the $\mathcal{O}(\alpha_{\rm s}^4)$ fixed-order prediction in the large transverse momentum region.

In Ref.~\cite{Becher:2012yn} using methods of SCET, the calculation of the cross sections for the Higgs boson production at small transverse momentum $q_{\rm T}$ region is performed, where large logarithms of the scale ratio $m_H/q_{\rm T}$ are resummed to all orders. The differential cross section based on SCET can be factorized as
\begin{align}\label{fact1}
   &\frac{{\rm d}^2\sigma}{{\rm d}q_{\rm T}^2\,{\rm d}y}
   = \sigma_0(\mu)\,C_t^2(m_t^2,\mu)
    \left| C_S(-m_H^2,\mu) \right|^2
    \nonumber\\
   &\quad\times \sum_{i,j=g,q,\bar q}\,
    \int_{\xi_1}^1\!\frac{{\rm d}z_1}{z_1} \int_{\xi_2}^1\!\frac{{\rm d}z_2}{z_2} \bar C_{gg\leftarrow ij}(z_1,z_2,q_{\rm T}^2,m_H^2,\mu)\, \nonumber \\
   &\quad\times
    \phi_{i/P}(\xi_1/z_1,\mu)\,\phi_{j/P}(\xi_2/z_2,\mu) \,,
\end{align}
where the Wilson coefficient $C_t$ can be obtained after integrating out the heavy top quark, while the hard matching coefficient $C_S$ arises when two-gluon operator in QCD is matched onto an effective two-gluon operator in SCET. $\phi_{i/P}$ is the ordinary parton distribution function (PDF). Besides, the integral kernel $\bar C_{gg\leftarrow ij}$ is
\begin{align}\label{Cdef}
   & \bar C_{gg\leftarrow ij}(z_1,z_2,q_{\rm T}^2,m_H^2,\mu)
   = \frac{1}{4\pi} \int\!{\rm d}^2x_\perp\,{\rm e}^{-{\rm i}q_\perp\cdot x_\perp} \nonumber\\
    &\quad\times \left( \frac{x_T^2 m_H^2}{b_0^2} \right)^{-F_{gg}(L_\perp,a_s)}
    \sum_{n=1,2}\,
    I_{g\leftarrow i}^{(n)}(z_1,L_\perp,a_s)\, \nonumber\\
    &\quad\times I_{g\leftarrow j}^{(n)}(z_2,L_\perp,a_s) \, ,
\end{align}
where $I_{g\leftarrow i}^{(n)}$ is the matching coefficient when matching the transverse momentum dependent PDF onto ordinary PDF. $F_{gg}$ is the collinear anomaly factor, which is first studied in Ref.~\cite{Becher:2010tm}. The results show that the resummation predictions are fully compatible with the NNLL order predictions~\cite{deFlorian:2011xf} obtained in the traditional CSS framework.

\begin{figure}[h]
  \centering
  \includegraphics[scale=0.7]{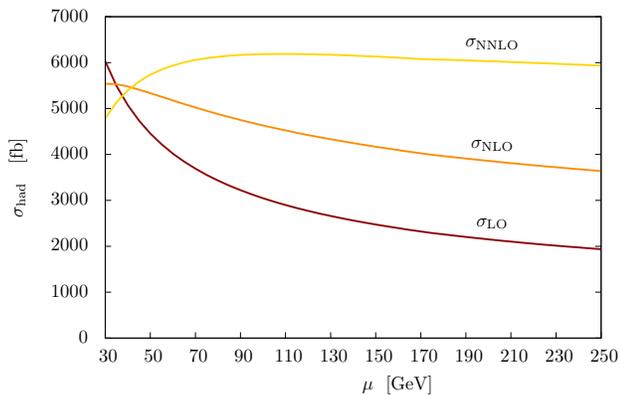} \hspace{0.5em}
  \vspace{-2ex}
  \caption{\label{fig:plot_scale} (Color online) Scale uncertainties for the process $gg\rightarrow h+1$ jet~\cite{Boughezal:2013uia} }
\end{figure}

Recently, the QCD NNLO corrections to the process $gg\rightarrow h+1~$jet were calculated~\cite{Boughezal:2013uia}. This is one of the first calculations, where QCD NNLO corrections are computed
for a $2\rightarrow2$ process, whose cross section depends on the implementation of the jet algorithm. The contribution to this process at $\mathcal{O}(\as^5)$ can be divided into three categories:
\begin{itemize}
\item  $gg \to H+g$ at two-loop level;
\item  $gg \to H+gg$ at one-loop level;
\item  $gg \to H+ggg$ at tree level.
\end{itemize}
In order to perform the complete QCD NNLO calculation, these three contributions have to be combined appropriately. In these calculations a key idea is utilized to deal with infrared divergences, which can be isolated through appropriate parameterizations of  phase-space and expansions in plus-distributions \cite{Frixione:1995ms}. To illustrate this method, consider the integral
\begin{align}
I(\epsilon) = \int \limits_{0}^{1} {\rm d} x x^{-1-a\epsilon} F(x),
\end{align}
where the  function $F(x)$ has a well-defined limit
$\lim \limits_{x \to 0}^{} F(x) = F(0)$. Expanding $I$ in $\epsilon$, the $x^{-1-a\epsilon}$ can be written as
\begin{align}
\frac{1}{x^{1+a\epsilon}} = -\frac{1}{a\epsilon} \delta(x) + \sum_{n=0}^{\infty}
\frac{(-\epsilon a)^n}{n!}\left[\frac{\ln^n(x)}{x}\right]_+,
\end{align}
so that
\begin{align}
I(\epsilon) & = \int \limits_{0}^{1} {\rm d} x \Big [ -\frac{F(0)}{a\epsilon}  + \frac{F(x) - F(0)}{x} \nonumber \\
 & -a\epsilon  \frac{F(x) - F(0)}{x} \ln(x) +...\Big ].
\end{align}
Here each term can be calculated independently. In Fig.~\ref{fig:plot_scale} the scale uncertainties for the process $gg\rightarrow h+1~$jet at LO, NLO and NNLO are shown, respectively. Obviously,  NNLO corrections increase NLO total cross section by about $30~\%$, and the scale uncertainties are reduced to less than $5\%$.

It is worth noting that based on the effective Lagrangian approximation with the form factor the processes $gg\rightarrow h+2~$jet and $gg\rightarrow h+3~$jet at the LO level were investigated~\cite{Campanario:2010mi,Campanario:2013mga}. Furthermore, NLO QCD corrections to these processes were also calculated recently~\cite{vanDeurzen:2013rv,Cullen:2013saa}.

\subsubsection{Vector boson fusion channel}

\begin{figure}[h]
  \centering
  \includegraphics[scale=0.5]{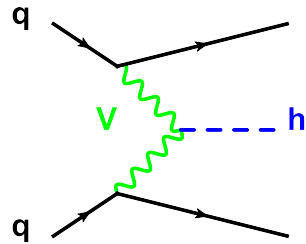} \hspace{0.5em}
  \vspace{-2ex}
  \caption{\label{fig:VBF} (Color online) Feynman diagram of $qq'\rightarrow jjh$ at the LO}
\end{figure}

At the LHC the SM Higgs boson can also be produced via VBF in association with two hard jets in the forward regions as shown in Fig~\ref{fig:VBF}. Through the VBF channel it is helpful to determine the couplings between Higgs and EW gauge bosons.

The QCD NLO corrections are of the order of $5~\%\sim10~\%$ and reduce the factorization and renormalization scale dependence of the cross section to a few percent~\cite{Spira:1997dg,Han:1992hr,Figy:2003nv,Figy:2004pt,Berger:2004pca}. The full EW NLO + QCD NLO corrections have been computed~\cite{Ciccolini:2007jr,Ciccolini:2007ec}, and the results show that the EW corrections are approximately 5~\%, which is as important as QCD corrections. Based on the structure function approach~\cite{Han:1992hr}, the approximate QCD NNLO corrections to the total cross section for VBF have been presented in Ref.~\cite{Bolzoni:2010xr}, and the scale uncertainty is reduced to $1\%\sim2~\%$ after combining QCD and EW calculations.

Recently, the theoretical predictions of VBF Higgs production plus one jet production are presented at QCD NLO level~\cite{Campanario:2013fsa}. The results show that the NLO corrections to the total cross section are moderate for the scale choice of $\mu=H_{\rm T} /2$, but can be more significant for $\mu=m_W /2$. Nevertheless, the scale uncertainty significantly decreases from around $30~\%$ ($24~\%$) at LO to about $2~\%$ ($9~\%$) at NLO, where the scale is chosen as $H_{\rm T}/2$($m_W/2$).

\subsubsection{Higgs strahlung production channel}
\begin{figure}[h]
  \centering
  \includegraphics[scale=0.5]{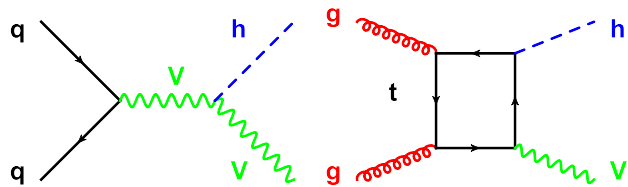} \hspace{0.5em}
  \vspace{-2ex}
  \caption{\label{fig:VH} (Color online) Feynman diagrams of $q\bar{q}\rightarrow hV$ and $gg\rightarrow hV$  at the lowest order}
\end{figure}

The associated production of Higgs boson $H$ and vector boson $V~(Z, W^{\pm})$ is the main channel of searching Higgs boson at the Tevetron, whose Feynamn daigrams are shown in Fig.~\ref{fig:VH}. However, by means of modern jet substructure methods, $HV$ production is also an important process to study the Higgs boson at the LHC. Two different decay modes, $h \rightarrow b\bar{b}$ and $h \rightarrow W^+ W^-$ have been searched by the ALTAS~\cite{ATLAS-CONF-2013-075,ATLAS-CONF-2013-079} and CMS~\cite{CMS-PAS-HIG-13-017,Chatrchyan:2013zna} collaborations.

The efforts of obtaining accurate theoretical predictions for $HV$ associated production at the hadron colliders have been made for a long time. The QCD NLO and EW corrections for $HV$ associated production were performed~\cite{Han:1991ia,Baer:1992vx,Ohnemus:1992bd,Kniehl:1990iva,
Ciccolini:2003jy}. Moreover, the QCD NNLO corrections of the total cross section for $HV$ associated production were calculated in Refs.~\cite{Hamberg:1990np,Harlander:2002wh,Brein:2003wg}. The corresponding numerical results are included in the code ${\rm VH}@{\rm NNLO}$~\cite{Brein:2012ne}.  Based on the transverse momentum substraction scheme\cite{Catani:2007vq}, the QCD NNLO corrections of differential cross section for $HW^\pm$ associated  production were completed~\cite{Ferrera:2011bk}. And the effects of QCD NLO corrections to both $HW^{\pm}$ associated production and subsequent decay of $H\to b\bar{b}$ were investigated~\cite{Banfi:2012jh}. Moreover, the fully differential cross section of this process up to QCD NNLO with the subsequent decay of the Higgs boson into $b\bar{b}$ at NLO is obtained~\cite{Ferrera:2013yga}.

\begin{figure}[h]
  \centering
  \includegraphics[scale=0.4]{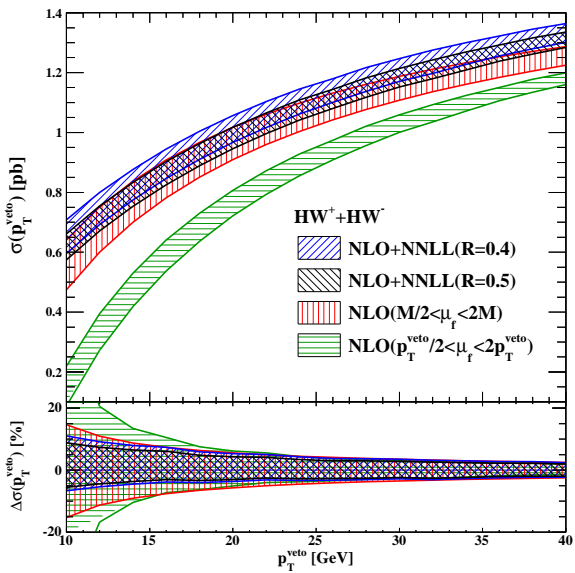} \hspace{0.5em}
  \vspace{-2ex}
  \caption{\label{fig:HW_RGI_vs_NLO}   The NLO+NNLL predictions for $HW$ associated production cross section with a jet veto at the 14~TeV LHC, where the bands reflect the scale uncertainties~\cite{Shao:2013uba}}
\end{figure}

Recently, the $HV$ associated production at the LHC with a jet veto is presented~\cite{Shao:2013uba}, where the large logarithmic terms ln$p_{\rm T}^{\rm veto}/Q$ existing in the perturbative expansions are resummed to all order in SCET. The resumed cross sections can be written as
\begin{align}\label{fac_cs_3}
 \frac{{\rm d}\sigma(p_{\rm T}^{\rm veto})}{{\rm d} M^2} = & \frac{\sigma_0}{s} \overline{H}(M,\ptv) \int_{\tau}^1 \frac{{\rm d} z}{z} \, \overline{\II}_{ij}(z, p_{\rm T}^{\rm veto},\mu_f) \nno \\
 & \times \ff_{ij}\left( \frac{\tau}{z},\mu_f \right),
\end{align}
where $\overline{\II}_{ij}$ and $\ff_{ij}$ are defined as
\begin{align}
 &\overline{\II}_{ij}(z, p_{\rm T}^{\rm veto},\mu_f) = \int_{z}^1\frac{{\rm d} u}{u}
 \overline{I}_{q \leftarrow i}(u,p_{\rm T}^{\rm veto},\mu_f)
  \nonumber \\ & \qquad \times \overline{I}_{\bar{q} \leftarrow j}(z/u,p_T^{\rm veto},\mu_f) + (q \leftrightarrow \bar{q} ),
 \\ &
 \ff_{ij}\left(y,\mu_f\right) = \int_y^1 \frac{{\rm d}x}{x}f_i(x,\mu_f)f_j(\frac{\tau}{x z},\mu_f).
\end{align}
Here $\overline{H}(M,\ptv)$ is the RG invariant hard function. Fig.~\ref{fig:HW_RGI_vs_NLO} shows the renormalization group improved NLO+NNLL predictions for $HW$ associated production cross section with a jet veto at the LHC. Obviously, the renormalization group improved predictions reduced the theoretical uncertainties, especially in small $\ptv$ region.

\subsubsection{Higgs boson associated with $t\bar{t}$ production\ }

\begin{figure}[h]
  \centering
  \includegraphics[scale=0.5]{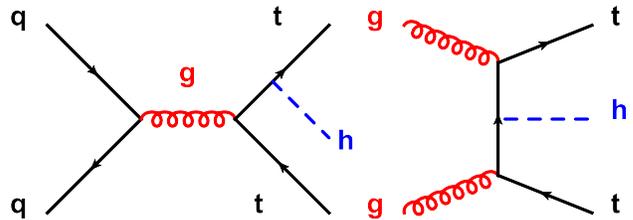} \hspace{0.5em}
  \vspace{-2ex}
  \caption{\label{fig:ttH} (Color online) Feynman diagrams of $q\bar{q}\rightarrow t\bar{t}h$ and $gg\rightarrow t\bar{t}h$  at the LO}
\end{figure}

The production of the Higgs boson associated with top quark pair, as shown in Fig.~\ref{fig:ttH}, is the main channel for measuring top quark and Higgs boson Yukawa coupling at the LHC. Similarly to top quark pair production, the LO predictions for $t\bar{t}h$ production suffer from large theoretical uncertainties~\cite{Raitio:1978pt,Ng:1983jm,Kunszt:1984ri,
Gunion:1991kg,Marciano:1991qq}. However, the QCD NLO results show~\cite{Beenakker:2001rj,Beenakker:2002nc,Reina:2001sf,
Dawson:2002tg} that QCD NLO corrections increase the total cross section by about $20~\%$, and the scale uncertainties are reduced to $10~\%$. Besides, recently QCD NLO corrections to Higgs boson production in association with $t\bar{t}+\rm{jet}$ were calculated~\cite{vanDeurzen:2013xla}.

\subsection{Higgs boson properties}

%\subsubsection{CP and spin}
In the SM, the Higgs boson is a $CP$-even, spin-0 particle ($J^P = 0^+$). The Landau-Yang theorem forbids the direct decay of a spin-1 particle into a pair of photons~\cite{landau:1948aa,yang:1950aa}. The spin-1 hypothesis is therefore strongly disfavored by the observation of the $h\rightarrow\gamma\gamma$ decay. The difference between the SM predictions $J^P = 0^+$ and alternative hypotheses can be studied through the bosonic decay channels $h\rightarrow\gamma\gamma$, $h\rightarrow WW^* \rightarrow 2l2\nu$ and $h\rightarrow ZZ^* \rightarrow 4l$,
 which recently are combined to distinguish between the SM assignment of $J^P = 0^+$ and a specific model of $J^P = 2^+$~\cite{ATLAS-CONF-2013-040}. Up to now the data strongly favor the $J^P = 0^+$ hypothesis, and the specific $J^P = 2^+$ hypothesis is excluded with a confidence level above $99.9~\%$, independently of the contributions of gluon fusion and quark-antiquark annihilation processes in the production of the spin-2 particle.

\subsubsection{Couplings}

\begin{figure}[h]
 \centering
  \includegraphics[scale=0.35]{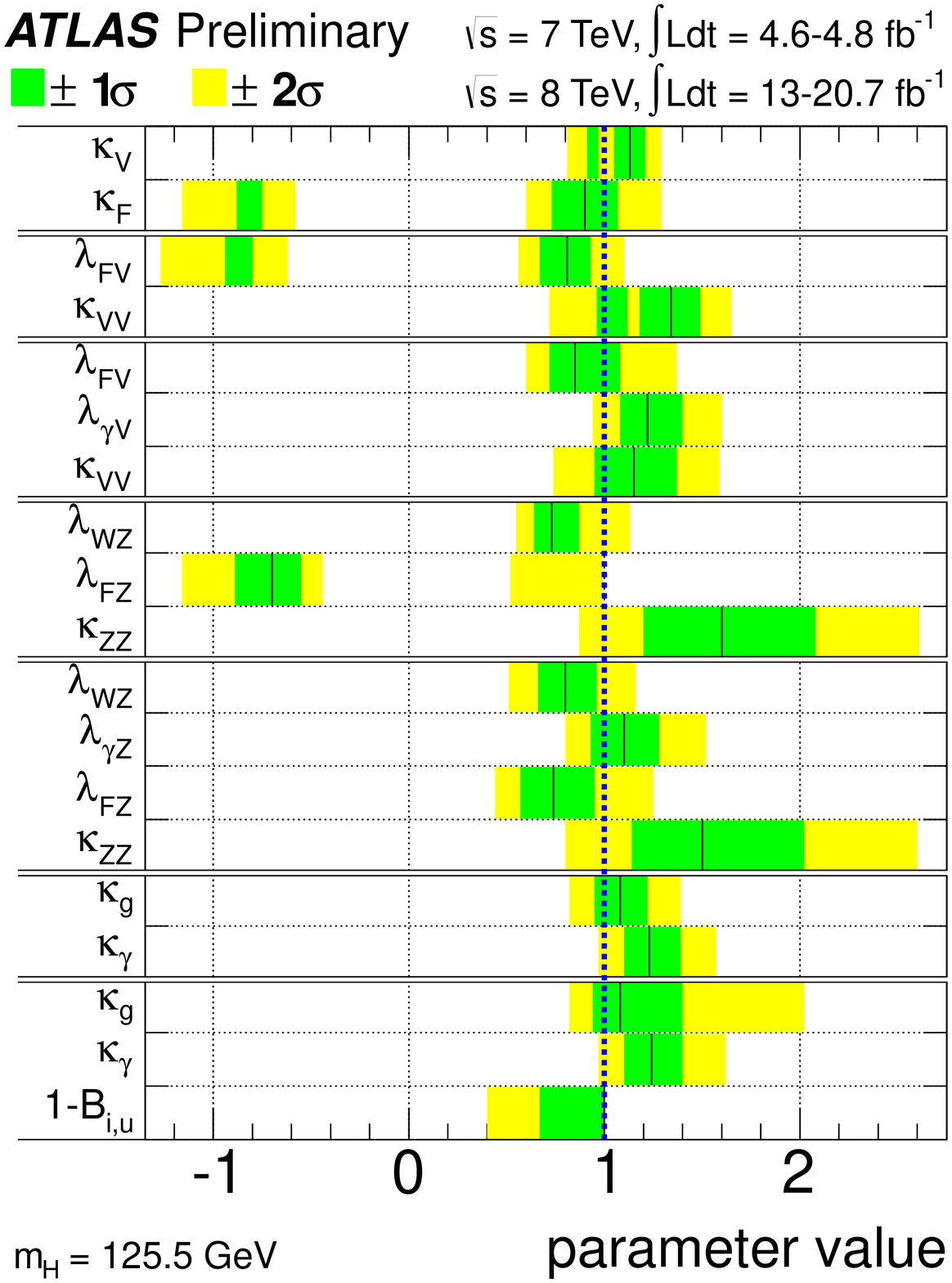}
  \quad
  \includegraphics[scale=0.35]{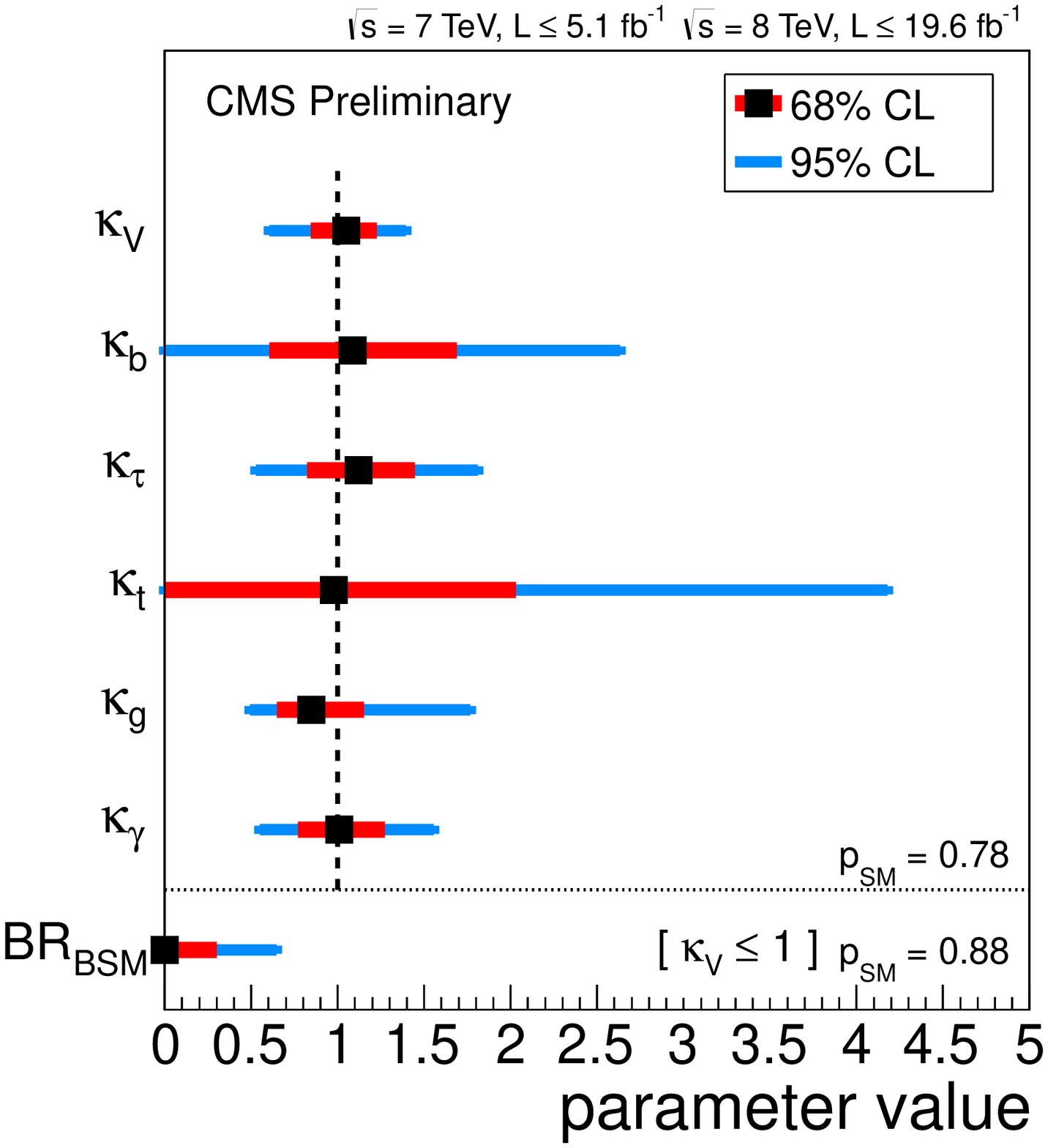}
  \caption{\label{fig:Higgs_coupling}   Summary of the coupling scale factor measurements for $m_H=125.5$ GeV~\cite{Aad:2013wqa,CMS-PAS-HIG-12-045} }
\end{figure}

Extraction of the Higgs coupling constants can serve to limit various new physics models, or further to confirm the validity of the SM. The deviations from the SM can be parameterized as scale factor $\kappa$ of Higgs couplings relative to the SM values:
\begin{eqnarray}
 g_{hff} = \kappa_f \cdot g_{hff}^{\rm SM}, ~~~ g_{hVV} = \kappa_V \cdot g_{hVV}^{\rm SM}.
\end{eqnarray}
Fig.~\ref{fig:Higgs_coupling} shows a summary of the coupling scale factor $\kappa$ measured by the ATLAS and CMS collaborations~\cite{Aad:2013wqa,CMS-PAS-HIG-12-045}, which indicates that the measured coupling between Higgs boson and other SM particle are consistent with the SM predictions. In the future, with the increasing of statistics of Higgs boson, the measurement of the couplings may be more precision.

\subsubsection{Self-coupling constant}
In the SM, the Higgs boson is responsible for the origin of EW symmetry breaking and the generation of elementary particle masses. After the Higgs field $\mathnormal{\Phi}$ gets the vacuum expectation value $v$, the SM Higgs potential in the unitary gauge can be written as
\begin{eqnarray}
 V(h)= \lambda \left[\frac{(v+h)^2}{2}-\frac{v^2}{2}\right]^2,
\end{eqnarray}
where the Higgs boson self-coupling $\lambda$ is given by $\lambda_{\rm SM}=m_H^2/(2v^2)$ at the tree-level in the SM, and the radiative corrections can decrease $\lambda_{\rm SM}$ by $10~\%$ for $m_{H}=125$ GeV where main contributions come from top quark loops~\cite{Kanemura:2002vm}.

\begin{figure}[h]
  \centering
  \includegraphics[scale=0.6]{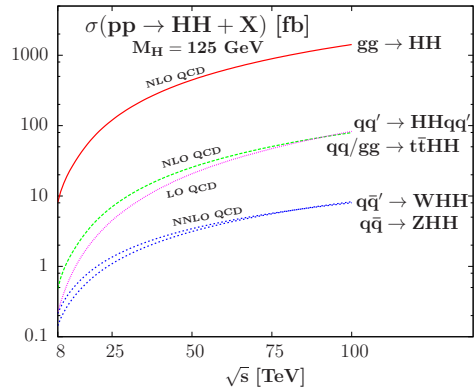} \hspace{0.5em}
  \vspace{-2ex}
  \caption{\label{fig:HH_main_fig}   The total cross sections for Higgs pair production at the LHC~\cite{Baglio:2012np}}
\end{figure}

At the LHC, the Higgs boson self-coupling $\lambda$ can be directly probed through Higgs boson pair production, and the relevant studies have been performed~\cite{Glover:1987nx,Plehn:1996wb,Dawson:1998py,
Djouadi:1999rca,Baglio:2012np,Binoth:2006ym,Baur:2002rb,Baur:2002qd,Shao:2013bz,
Baur:2003gpa,Baur:2003gp,Dolan:2012rv,Papaefstathiou:2012qe,
deFlorian:2013uza,Gupta:2013zza,Yao:2013ika,Barr:2013tda,deFlorian:2013jea,
Dolan:2013rja,Li:2013flc,Goertz:2013kp,Grigo:2013rya,Grigo:2013xya}. Similarly to the case of single Higgs boson production, there are four classes of Higgs pair production at the LHC, and the corresponding total cross sections are shown in Fig.~\ref{fig:HH_main_fig} as functions of the center of mass frame energy.

\begin{figure}[h]
  \centering
  \includegraphics[scale=0.5]{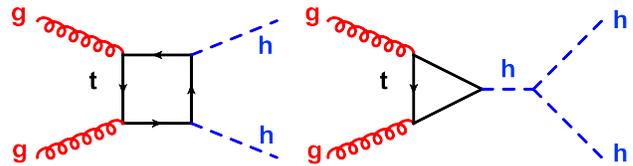} \hspace{0.5em}
  \vspace{-2ex}
  \caption{\label{fig:gg2HH} (Color online) Feynman diagrams of $gg \rightarrow hh$ at the lowest order}
\end{figure}

The Higgs boson pair production is mainly induced by gluon gluon fusion (see Fig.~\ref{fig:gg2HH}). In Ref.~\cite{Dawson:1998py}, the QCD NLO corrections are calculated in the large top quark mass limit. Recently, the soft gluon threshold resummation and $\pi^2$ enhancement effects in Higgs boson pair production at the LHC have been calculated in Ref.~\cite{Shao:2013bz}. In the infinite top quark mass limit, the effective Lagrangian describing $ggh$ and $gghh$ interactions is given by
\begin{align}\label{effL}
 \mathcal{L}_{\rm eff}=&\frac{\as(\mu^2)}{12\pi v}C_t(\mu^2)G_{\mu\nu}^a G^{a~\mu\nu}h \nno \\
 &- \frac{\as(\mu^2)}{24\pi v^2}C_t(\mu^2)G_{\mu\nu}^a G^{a~\mu\nu}h^2.
\end{align}
Up to NLO, the Wilson coefficient $C_t(\mu^2)$ was calculated by performing the large top quark mass expansion of the  corresponding one- and two-loop Feynman diagrams~\cite{Dawson:1998py}. In SCET the differential cross section can be factorized as
\begin{align}
\label{cs}
\frac{{\rm d}^3\sigma}{{\rm d}M^2{\rm d}Y{\rm d}\cos\theta} & = \frac{\as^2 G_{\rm F}^2 M^2 \beta_H}{2304(2\pi)^3S}  \Bigg[ \Big|f_{\rm Tri}^{\rm A} + f_{\rm Box}^{\rm A}\Big|^2 + \Big| f_{\rm Box}^{\rm B} \Big|^2 \Bigg] \nno\\
 &\hspace{-4em} \times   \Bigg[ \int_{\sqrt{\tau} \text{e}^{-Y}}^{1}\frac{{\rm d}z}{z}f_{g/A}(\sqrt{\tau}\text{e}^{Y},\mu_f)
 f_{g/B}(\frac{\sqrt{\tau}}{z}\text{e}^{-Y},\mu_f) \nno \\
 & \hspace{-4em}  +\int_{\sqrt{\tau} \text{e}^{Y}}^{1}\frac{{\rm d}z}{z}f_{g/A}(\frac{\sqrt{\tau}}{z}\text{e}^{Y},\mu_f)
 f_{g/B}(\sqrt{\tau}\text{e}^{-Y},\mu_f) \Bigg] \nno \\
 &\hspace{-4em} \times\frac{C(z,M,\mu_f)}{2}.
\end{align}
where $f_{\rm Tri}^{\rm A}, f_{\rm Box}^{\rm A}$ and $f_{\rm Box}^{\rm B}$ are the form factors including complete top quark effects at one-loop level. The integral kernel $C(z,M,\mu_f)$ has the form
\begin{align}\label{Cresum}
 C(z,m_t,M,\mu_f) & = \left[C_t(m_t^2,\mu_t^2)\right]^2 \left|C_S(-M^2,\mu_h^2)\right|^2 \nno\\
 & \hspace{-4em} \times U(M^2,\mu_t^2,\mu_h^2,\mu_s^2,\mu_f^2) \frac{z^{-\eta}}{(1-z)^{1-2\eta}} \nno\\
 & \hspace{-4em} \times \tilde{s}\left(\ln\frac{M^2(1-z)^2}{\mu_s^2z} +\partial_\eta,\mu_s^2\right)\frac{\text{e}^{-2\gamma_{\rm E}\eta}}{\mathnormal{\Gamma}(2\eta)},
\end{align}
where $C_S$ is the hard matching coefficient, $\tilde{s}$ is the soft function, and $U$ is the evolution function.
\begin{table}
\vspace{0.2cm}
\centering
\caption{\label{lam_totxsec_1}
NLO and NLO+NNLL total cross sections of Higgs boson pair production at the 14~TeV LHC for different Higgs boson self-coupling $\lambda$. The first errors represent the scale uncertainties. The second errors are PDF+$\as$ uncertainties~\cite{Shao:2013bz}. }
\begin{tabular}{cccc}
\hline
\multirow{2}{*}{$\lambda/\lambda_{\rm SM}$} & \multicolumn{3}{c}{$\sqrt{S}=14~{\rm TeV}$} \\[-0.1cm]
  & NLO~(fb) & NLO + NNLL~(fb) & $K$-factor \\
\hline
\\[-0.4cm]
$-1$ &
 $127.9^{+23.1+8.7\,(+3.8)}_{-18.8-7.7\,(-3.3)}$ &
 $161.6^{+9.8+12.0\,(+6.0)}_{-3.1-11.4\,(-4.9)}$ &
 1.26
\vspace{0.2cm}\\
0 &
 $71.1^{+12.8+4.8 \, (+2.1)}_{-10.5 - 4.3 \, (-1.8)}$ &
 $90.0^{+5.4+6.8 \, (+3.3)}_{-1.7 - 6.4 \, (-2.8)}$ &
 1.27
\vspace{0.2cm}\\
1 &
 $33.9^{+6.1 + 2.3 \, (+1.0)}_{-5.0 - 2.0 \, (-0.9)}$ &
 $42.9^{+2.6 + 3.3 \, (+1.6)}_{-0.8 - 3.1 \, (-1.3)}$ &
 1.27
\vspace{0.2cm}\\
2 &
 $16.1^{+2.9+1.1 \, (+0.5)}_{-2.4 - 1.0 \, (-0.4)}$ &
 $20.4^{+1.2+1.6 \, (+0.8)}_{-0.4 - 1.5 \, (-0.7)}$ &
 1.27
\vspace{0.2cm}\\\hline
\end{tabular}

\end{table}

\begin{figure}[h!]
  \centering
  \includegraphics[scale=0.4]{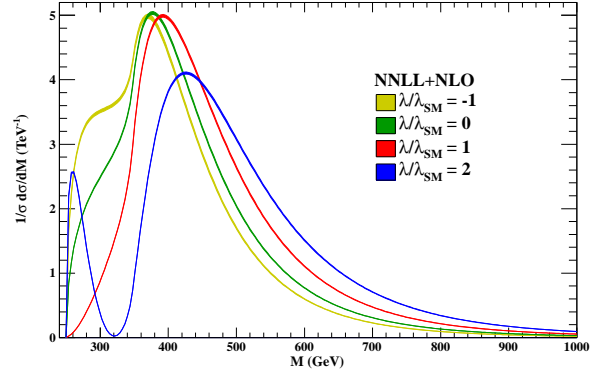} \hspace{0.5em}
  \vspace{-2ex}
  \caption{\label{fig:lam_mhh_14} The normalized Higgs boson pair invariant mass distribution at the LHC with $\sqrt{S}=14$ TeV, where the bands represent the scale uncertainties~\cite{Shao:2013bz}}
\end{figure}

Table~\ref{lam_totxsec_1} shows the NLO and NLO+NNLL total cross sections of Higgs boson pair production at the LHC with $\sqrt{S}=14$~TeV for different Higgs boson self-coupling $\lambda$. Obviously, due to the interference effects between two channels, the total cross section of Higgs boson pair production decreases with the increasing of $\lambda$. Besides, the resummation effects increase the QCD NLO results by about $20~\% - 30~\%$. Moreover, in Fig.~\ref{fig:lam_mhh_14} the resummation results show that the shape of the normalized invariant mass distribution of Higgs boson pair strongly depends on the Higgs boson self-coupling $\lambda$. And it is possible to extract the parameter $\lambda$ from the Higgs boson pair invariant mass distribution.

Later, the extraction of the Higgs self-coupling is also studied by exploiting the double-to-single cross section ratio~\cite{Goertz:2013kp}. The top quark mass effects on the total cross section of Higgs boson pair production at the QCD NLO have been studied~\cite{Grigo:2013rya,Grigo:2013xya}, where the NLO cross section keeping exact top quark mass is expanded in powers of $1/m_t$. And the the power corrections are calculated up to $O(1/m_t^8)$ and $O(1/m_t^{12})$ for partonic channel $gg \to HH$ and $qg(\bar{q}) \to HH$, respectively. They find that the poor convergence induced by top quark mass effects can be improved if the exact LO cross section are used to normalize the QCD NLO correction, and the remaining uncertainties from top mass effects are about $\mathcal{O}(10~\%)$ in the QCD NLO results.

 Very recently, the full QCD NNLO corrections for the cross section in the large top mass limit for Higgs boson pair production are calculated, in which the soft and collinear divergences are removed via the FKS subtraction method~\cite{deFlorian:2013jea}. Table~\ref{tabla} shows the total cross section as functions of the center of mass frame energy at NNLO.

\begin{table}[h]
\begin{center}
\caption{Total cross section as functions of the center of mass frame energy at NNLO accuracy. The exact LO prediction to normalize the results is used~\cite{deFlorian:2013jea}.}
\begin{tabular}{l  c  c  c  c }
\hline
$E_{\rm cm}$ & $8\text{ TeV}$ & $14\text{ TeV}$ & $33\text{ TeV}$ & $100\text{ TeV}$ \\
\hline
$\sigma_{\text{NNLO}}$ & $9.76\text{ fb}$ & $40.2\text{ fb}$ & $243\text{ fb}$ & $1638\text{ fb}$ \\
Scale $(\%)$ &\footnotesize $+9.0-9.8\,$ &\footnotesize $\,+8.0-8.7\,$ &\footnotesize $\,+7.0-7.4\,$ &\footnotesize $\,+5.9-5.8$ \\
PDF $(\%)$ &\footnotesize $+6.0-6.1\,$ &\footnotesize $\,+4.0-4.0\,$ &\footnotesize $\,+2.5-2.6\,$ &\footnotesize $\,+2.3-2.6$ \\
PDF$+\as$  $(\%)$ &\footnotesize $+9.3-8.8\,$ &\footnotesize $\,+7.2-7.1\,$ &\footnotesize $\,+6.0-6.0\,$ &\footnotesize $\,+5.8-6.0$ \\
\hline
\end{tabular}
\label{tabla}
\end{center}
\end{table}

\section{Recent progress in top quark physics}

Due to the large mass of top quark, it is one of the hottest topics in particle physics. Top quarks are mostly produced through top and anti-top pairs production via strong interactions, or single top production via EW interactions at hadron colliders.

\subsection{Top quark mass determination}

\begin{figure}[t!]
  \centering
  \includegraphics[scale=0.40]{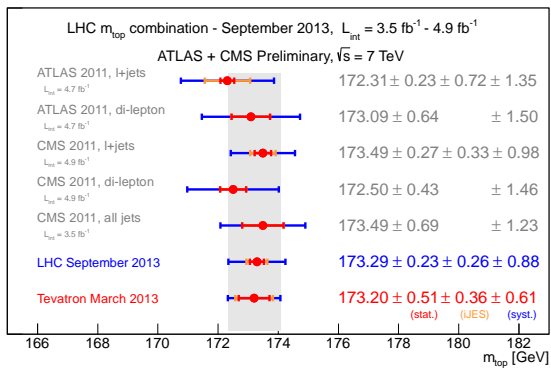}
  \vspace{-2ex}
  \caption{\label{fig:top_mass}
 (Color online) The combined measurements from the LHC compared with the Tevatron combined top mass measurements~\cite{CDF:2013jga,TheATLAScollaboration:2013hja}}
\end{figure}

The top quark mass is one of the free parameters in SM. Through EW corrections, the top quark mass together with the $W$ boson mass can be used to constrain the Higgs boson mass. Thus, the precise top quark mass is important for testing the SM or searching for NP, using precision EW fits. It was pointed out in Ref.~\cite{Agashe:2013hma} that precise measurement of $M_W$ in future requires the high precision top quark mass so that EW precision fits are not restricted by the uncertainty of the top quark mass~\cite{Baak:2012kk,Baak:2013fwa}. Top quark mass also plays a crucial role in constraining Higgs boson mass by vacuum stability of Higgs field. If changing top quark mass by 2.1 GeV around the central value  $m_t = 173.1$ GeV, vacuum instability scale changes from $\mu_{\rm neg} \sim 10^{8} $ GeV to $\mu_{\rm neg} \sim 10^{14} $ GeV~\cite{Degrassi:2012ry,Agashe:2013hma}. Thus, the precision determination of  top quark mass has an important impact on the understanding of the SM.

Because top quark mass can not be measured directly, it can be only extracted from observables which is sensitive to it. In the following, we briefly review some popular methods.

 The matrix element method~\cite{EstradaVigil:2001eq,Abazov:2004cs} used at the Tevatron is the most precise approach for the measurements of the top quark mass, where the measured results are compared with predictions of the LO $t\bar{t}$ production and decay convoluted with the detector response. An approach including NLO QCD effects is being developed~\cite{Campbell:2012cz}. Another most precise approach is ideogram and template methods which are used by the ATLAS and CMS. With these approaches, the top quark mass is determined by comparing the reconstructed distributions with Monte Carlo spectrums. A third approach is  extracting the top quark mass from the total cross section of top-pair production, with which the latest results was performed at the NNLO+NNLL level~\cite{Czakon:2013goa}. However, the sensitivity of total cross section to $m_t$ is relatively small, which means that with this method  the precision is lower than the ones of others. Besides above methods, there are other approaches for the
determination of $m_t$, which are reviewed  recently in Refs.~\cite{Mitov:2013jla,Juste:2013dsa}.

In Fig.~\ref{fig:top_mass}, the combined measurements from the LHC are compared with the latest ones from the Tevatron~\cite{TheATLAScollaboration:2013hja,CDF:2013jga}, which shows that the combined results are $m_t = (173.2 \pm 0.87)\  {\rm{GeV}}$ at the Tevatron and $m_t = (173.3 \pm 0.95)\ {\rm{GeV}}$ at the LHC.

\subsection{Top quark decay at NNLO}

The top quark decay width has already been directly measured at the Tevatron~\cite{Aaltonen:2013kna}. In SM  the top quark almost 100~\% decays into a bottom quark and a $W$ boson. The QCD NLO calculations were done over twenty years ago~\cite{Jezabek:1988iv,Czarnecki:1990kv,Li:1990qf}.
The EW corrections were computed at NLO accuracy~\cite{Eilam:1991iz,Denner:1990ns}. In the approximation of $m_t \gg m_W$, the QCD NNLO corrections to width were calculated \cite{Czarnecki:1998qc}. Based on the calculations of top quark self-energy as an expansion in $m_W^2/m_t^2$, the NNLO decay width was presented in Refs.~\cite{Chetyrkin:1999ju,Blokland:2004ye}.

Recently, the calculation of top quark decay width at NNLO in QCD, including NLO EW corrections as well as the finite bottom quark mass and $W$ boson width effects, was completed in Ref.~\cite{Gao:2012ja}, where the NNLO fully differential decay rates were first presented. Later, another calculation of NNLO differential top quark decay width was presented in Ref.~\cite{Brucherseifer:2013iv}.

\begin{table}[h!]
\caption{\label{ttop} Top quark total width including the QCD NLO and NNLO corrections and NLO EW corrections~\cite{Gao:2012ja}.}
  \begin{center}
      \begin{tabular}{c c ccccc}
        \hline
          $m_t~(\rm{GeV})$ & $\mathnormal{\Gamma}_t^{(0)}~(\rm{GeV})$ & $\delta_f^b$ & $\delta^W_f$ & $\delta_{\rm EW}$ &
          $\delta^{(1)}_{\rm QCD}$ & $\delta^{(2)}_{\rm QCD}$ \\
        \hline
        172.5 &1.4806 & $-0.26$ & $-1.49$ &1.68 & $-8.58$ & $-2.09$\\
                173.5 &1.5109 & $-0.26$ & $-1.49$ &1.69 & $-8.58$ & $-2.09$\\
              174.5 &1.5415 & $-0.25$ & $-1.48$ &1.69 & $-8.58$ & $-2.09$\\
        \hline
      \end{tabular}
  \end{center}
  \vspace{-3ex}

\end{table}

We briefly review the method proposed in Ref.~\cite{Gao:2012ja}. In the NLO and NNLO calculations, the bottom quark mass is set as $m_b=0$. All the partons in the final state are clustered into a single jet whose invariance mass is measured by $\tau=(p_b+p_X)^2/m_t^2$. Therefore,  when $\tau\to 0$ the radiations can only be soft and(or) collinear to bottom quark. We can divide the top quark decay width into two parts:
\begin{equation}
\label{eq:gamma}
\mathnormal{\Gamma}_t = \int^{\tau_0}_0\!\mathrm{d}
\tau\,\frac{\mathrm{d}\mathnormal{\Gamma}_t}{\mathrm{d} \tau} + \int^{\tau_{\rm max}}_{\tau_0} \!\mathrm{d}\tau
\frac{\mathrm{d}\mathnormal{\Gamma}_t}{\mathrm{d} \tau} \equiv \mathnormal{\Gamma}_A + \mathnormal{\Gamma}_B,
\end{equation}
where $\tau_0$ is a dimensionless cutoff for $\tau$ and $\tau_{\rm max}=(1-m_W/m_t)^2$. In the limit of $\tau \to 0$, ${\rm d}\mathnormal{\Gamma}_t/{\rm d}\tau$ can be expressed as
\begin{eqnarray}
\frac{1}{\mathnormal{\Gamma}^{(0)}_t}\frac{\mathrm{d} \mathnormal{\Gamma}_t}{\mathrm{d} \tau}
&=& \mathcal{H}\left( x\equiv \frac{m^2_W}{m^2_t},\mu\right) \int\!
\mathrm{d} k \, \mathrm{d} m^2 J(m^2,\mu) S(k,\mu) \nonumber \\ &&\times\delta \left( \tau
- \frac{m^2 + 2 E_J k}{m^2_t} \right) + \cdots.
\label{eq:fac}
\end{eqnarray}
The second part in Eq.~(\ref{eq:gamma}) can be obtained from  the QCD NLO corrections to $t\to W^+ b + \rm{jet}$.

\begin{figure}[h]
  \begin{center}
    \includegraphics[width=0.52\textwidth]{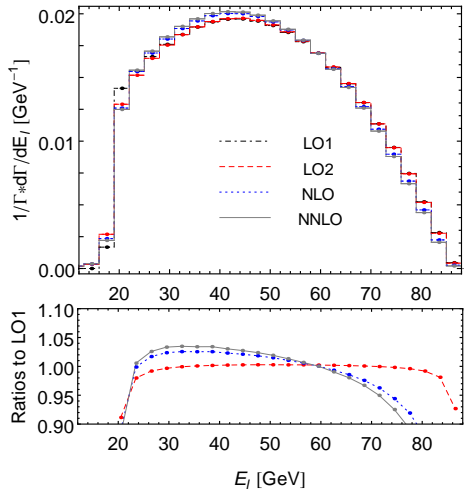}
  \end{center}
  \vspace{-5ex}
  \caption{\label{fig:el}
  Distribution of the charged lepton energy in top quark  rest frame~\cite{Gao:2012ja}}
\end{figure}

The total width is independent of $\tau_0$ as long as $\tau_0$ is sufficient small, which can be written as
\begin{eqnarray}
         \mathnormal{\Gamma}_t = \mathnormal{\Gamma}_t^{(0)} (1+\delta_f^b + \delta_f^W + \delta_{\rm EW} + \delta_{\rm QCD}^{(1)}+ \delta_{\rm QCD}^{(2)}),
\end{eqnarray}
where $\mathnormal{\Gamma}_t^{(0)}$ is the leading order total width, $\delta_f^b$ and $\delta_f^W$ is the effects of the finite $b$ quark mass and $W$ boson width, $\delta_{\rm EW}$ is the NLO EW corrections, and $\delta_{\rm QCD}^{(1)}$ and $\delta_{\rm QCD}^{(2)}$ are the QCD NLO and NNLO corrections, respectively. All these corrections  are shown in Table \ref{ttop}. After including the QCD NNLO corrections, the scale dependence is reduced to about $0.8~\%$, which makes the predictions more reliable. Fig.~\ref{fig:el} shows the charged lepton energy distribution.  It can be seen that the high order corrections push the energy distribution  into the central region.

The QCD NNLO corrections to top quark decay rates complement the QCD NNLO predictions for top quark pair production. This method can be also used in studies of heavy-to-light quark decay, such as B meson semileptonic decay.

\subsection{Top quark pair production}

The main channels of top-pair production at hadron colliders are
\begin{eqnarray}
      q(p_1) + \bar{q} (p_2) &\to& t(p_3)+\bar{t}(p_4),
      \nonumber \\
      g(p_1) + g (p_2) &\to& t(p_3)+\bar{t}(p_4).
\end{eqnarray}
For later convenience, we define the kinematic variables
\begin{eqnarray}
     \hat{s} = (p_1 +p_2)^2 \ , &&   \qquad  M_{t\bar{t}} = (p_3+p_4)^2\ ,  \nonumber \\
     t_1 = (p_1-p_3)^2-m_t^2\ , && \qquad u_1 = (p_1-p_4)^2-m_t^2\ .
\end{eqnarray}
And we use $p_{\rm T}$ and $q_{\rm T}$ to denote the transverse moment of the top quark and $t\bar{t}$ system, respectively.

\subsubsection{ Forward-backward asymmetry}

In the SM, the forward-backward asymmetry of top quark pair production at $p\bar{p}$ colliders mainly arises from the higher orders corrections in perturbative QCD~\cite{Kuhn:1998jr,Kuhn:1998kw}. As a result of it, the top quark is preferably produced in the direction of the incoming quark, and the anti-top follows the direction of incoming anti-quark. Thus, we can define the forward-backward asymmetry as
\begin{eqnarray}
       A_{\rm FB} = \frac{N(\Delta y > 0) - N(\Delta y < 0)} {N(\Delta y > 0) + N(\Delta y < 0)},
\end{eqnarray}
where $\Delta y  = y_t - y_{\bar{t}} $ is a difference of the rapidity of top and anti-top quark. The differences between the SM prediction and the latest measurements by the D0 and CDF are around 2$\sigma$~\cite{Abazov:2011rq,Aaltonen:2012it}. For the differential asymmetries, $A_{\rm FB}$ is found to have strong dependence on $t\bar{t}$ invariant mass, rapidity difference of top and anti-top quark, and the transverse momentum of the $t\bar{t}$ system. And the dependencies of the differential asymmetries on $|\Delta y | $ and $M_{t \bar{t}}$ show large deviation  by about 3$\sigma$ from the predictions of the SM~\cite{Abazov:2011rq,Aaltonen:2012it}.

The EW corrections were found to enlarge the SM predictions~\cite{Hollik:2011ps} and the interferences between the EW and QCD corrections also contribute to the asymmetry~\cite{Bernreuther:2012sx}. Beyond QCD NLO effects, the soft-gluon resummation predictions for $A_{FB}$ were presented in Refs.~\cite{Almeida:2008ug,Ahrens:2011uf}. It was shown that the soft gluon resummation corrections are very small and increase the NLO total asymmetry by less than $3~\%$. Recently, it was pointed out~\cite{Brodsky:2012rj} that the SM predictions actually have only 1 $\sigma$ deviation in the large pair invariant mass region from the CDF and D0 measurements  after the principle of maximum conformality scale setting~\cite{Brodsky:2011ig}. However, the calculations of a combination of $t\bar{t}$ and $t\bar{t}+\rm{jet}$  at QCD NLO merged with parton shower show that the dependences of the asymmetries on the rapidity and invariant mass still suggest a $2\sigma$ deviation from the experimental measurements~\cite{Hoeche:2013mua}.

At the LHC, it is difficult to measure the asymmetric, because the proton-proton collisions are forward-backward symmetry and a large gluon flux reduces the asymmetry here. But the charge asymmetry at the LHC can be measured through the difference in the top and anti-top rapidity distributions, which is defined as
\begin{eqnarray}
       A_{\rm C} = \frac{N(\Delta |y| > 0) - N(\Delta |y| < 0)}{N(\Delta |y| > 0) + N(\Delta |y| < 0)},
\end{eqnarray}
where $\Delta | y | = | y_t | - | y_{\bar{t}} | $. The measurements at the LHC have been performed by the ATLAS and CMS, which are found to be consistent with SM predictions~\cite{Chatrchyan:2011hk, ATLAS:2012an,Aad:2013cea,CMS:2013nfa}. Due to the large errors, these results are not conclusive. With sufficient luminosity,  the 14 TeV LHC will have enough sensitivity to the charge asymmetry~\cite{Jung:2013vpa}.

\subsubsection{Cross section at fixed-order}

\begin{figure*}
  \centering
  \includegraphics[width=0.4\textwidth]{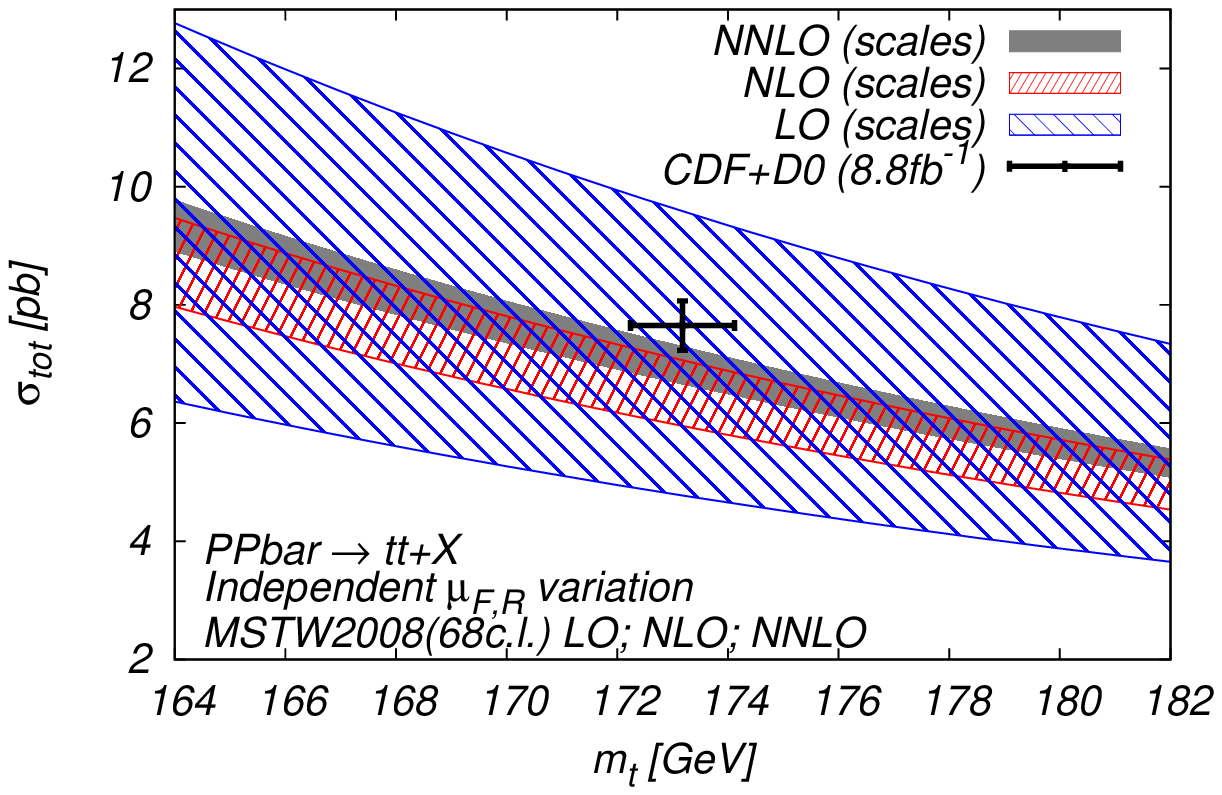}
  \includegraphics[width=0.4\textwidth]{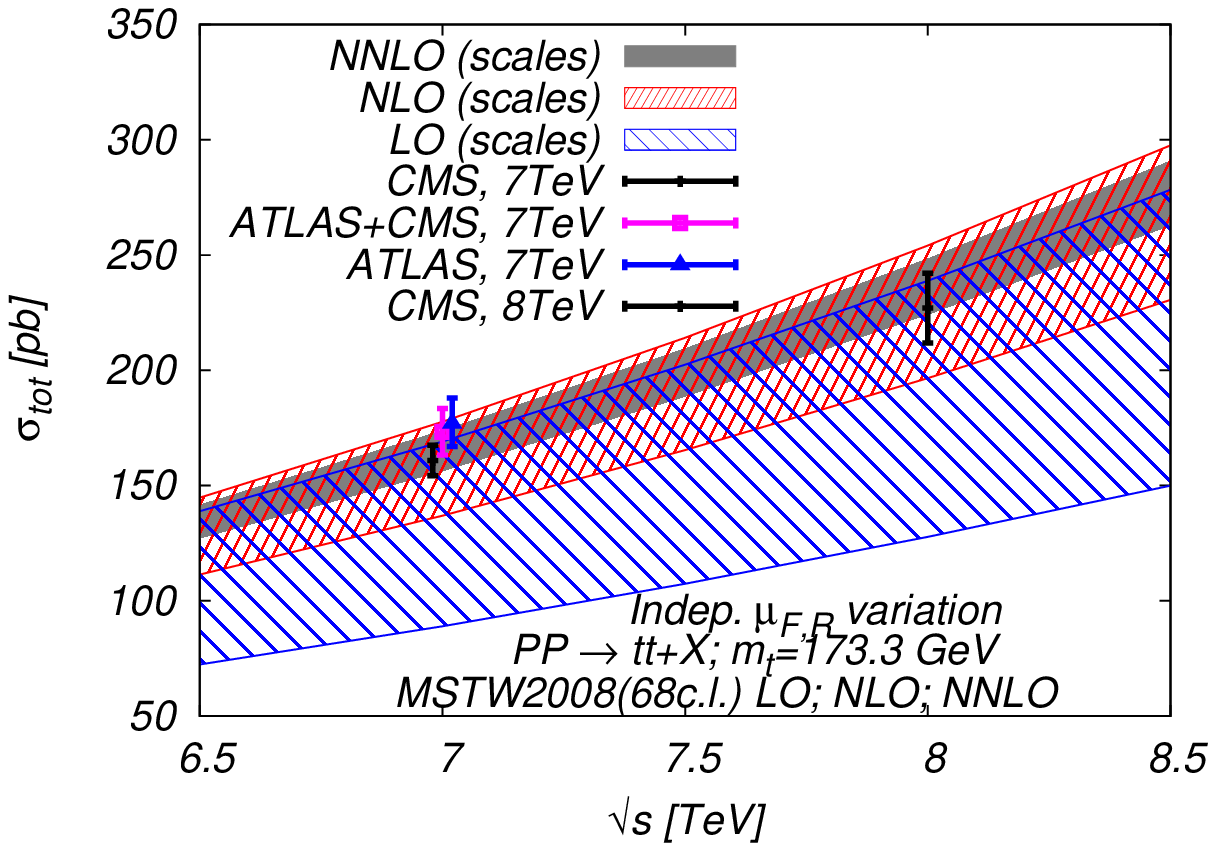}
  \vspace{-2ex}
  \caption{\label{fig:nnlo_scale} Scale dependence of the total  $t\bar{t}$ production cross section at the Tevatron and the LHC~\cite{Czakon:2013xaa}}
\end{figure*}

The total and differential cross sections for $t\bar{t}$ production were calculated at QCD NLO level over twenty years ago~\cite{Nason:1987xz, Nason:1989zy,Beenakker:1988bq,Beenakker:1990maa}. The EW radiative corrections were also calculated~\cite{Beenakker:1993yr,Bernreuther:2005is,Moretti:2006nf,Kuhn:2006vh}. And the QCD NLO corrections to top-pair production and decay at hadron colliders were obtained in narrow width approximation~\cite{Melnikov:2009dn,Bernreuther:2010ny,Campbell:2012uf}. Furthermore, the off-shell effects for top quark pair production were also studied at NLO~\cite{Denner:2010jp,Denner:2012yc,Bevilacqua:2010qb,Falgari:2013gwa}.
Very recently, it is extended to the case of massive $b$-quarks~\cite{Cascioli:2013wga}.

\begin{table}[h]
\begin{center}
\caption{\small The NNLO theoretical predictions for top-pair production at the Tevatron and the LHC~\cite{Czakon:2013goa}.}
\begin{tabular}{c c c c}
\hline
Collider & $\sigma_{\rm tot}$ [pb]  & scales [pb] & pdf [pb] \\
\hline
Tevatron & $7.009$ & $ ^{+0.259 (3.7\%)}_{-0.374 (5.3\%)} $ & $ ^{+0.169 (2.4\%)}_{-0.121 (1.7\%)} $   \\
LHC 7 TeV & $167.0$ & $ ^{+6.7 (4.0\%)}_{-10.7 (6.4\%)} $ & $ ^{+4.6 (2.8\%)}_{-4.7 (2.8\%)} $  \\
LHC 8 TeV & $239.1$ & $ ^{+9.2 (3.9\%)}_{-14.8 (6.2\%)} $ & $ ^{+6.1 (2.5\%)}_{-6.2 (2.6\%)} $  \\
LHC 14 TeV & $933.0$ & $ ^{+31.8 (3.4\%)}_{-51.0 (5.5\%)} $ & $ ^{+16.1 (1.7\%)}_{-17.6 (1.9\%)} $   \\
\hline
\end{tabular}

\label{tab:NNLO}
\end{center}
\end{table}

\begin{widetext}
\begin{center}
\begin{table}[h!]
\caption{\label{tab:soft_thresh} Three cases in soft gluon resummation for top-pair production.}
      \begin{tabular}{c c c c}
        \hline
          Name & Soft limit & Logarithmic corrections & Observables \\
          \hline
          Production threshold &     $\beta=\sqrt{1- \frac{4 m_t^2}{\hat{s}}}\to 0 $ & $ \frac{\ln^m \beta}{\beta^n} $ & Total cross section $\sigma$ \\
          Top pair invariant mass (PIM)  & $1-\frac{M_{t\bar{t}}^2}{\hat{s}} \to 0 $  & $ \frac{\ln^n(1-z)}{(1-z)} $ & $\frac{d\sigma} {dM_{t\bar{t}} d\cos(\theta) }$ \\
          Single particle inclusive (1PI) &  $s_4= \hat{s}+t_1+u_1 \to 0 $  &  $ \frac{\ln^m (s_4/m_t^2)}{s_4} $  & $ \frac{d\sigma}{dp_T dy}$\\               \hline
      \end{tabular}
\end{table}
\end{center}
\end{widetext}

The QCD NNLO calculations can be divided into three parts, i.e.  two-loop diagrams or one loop squared contributions without additional partons emitting, one-loop diagrams with one additional parton in the final state and tree level diagrams with emitting two additional partons. One-loop squared contributions were calculated in Refs.~\cite{Korner:2008bn,Anastasiou:2008vd,Kniehl:2008fd}. The analytic results for the two loop were calculated in the high energy limit~\cite{Czakon:2007ej,Czakon:2007wk}. And then, the analytic leading color contributions for the $q\bar{q}$ and $gg$ channel were computed~\cite{Bonciani:2008az,Bonciani:2009nb,Bonciani:2010mn}. Recently the exact results for the two loop contributions  were numerically calculated~\cite{Czakon:2008zk,Baernreuther:2013caa}. The contributions from the one-loop diagram with one additional parton emitting can be obtained through the NLO corrections of $t\bar{t}+\rm{jet}$~\cite{Dittmaier:2007wz,Dittmaier:2008yq,Bevilacqua:2010ve,Melnikov:2010iu}. The
double real radiation has been calculated with different subtraction method in Refs.~\cite{Czakon:2010td,Czakon:2011ve,Anastasiou:2010pw,Abelof:2011jv,Bernreuther:2011jt}. Based on the above progress, the NNLO total cross section  for top quark pair production has been completed~\cite{Baernreuther:2012ws,Czakon:2012zr,Czakon:2012pz,Czakon:2013goa}. The total cross sections for top-pair production are shown in Table~\ref{tab:NNLO}~\cite{Czakon:2013goa}, where the scale uncertainty is about 4~\% and 5~\% at the Tevatron and the LHC, respectively. Fig.~\ref{fig:nnlo_scale} shows that the scale dependence of the NNLO total cross sections is much smaller than that of the LO and NLO cross sections~\cite{Czakon:2013xaa}. It can be seen in Fig.~\ref{fig:nnlo_scale} that, at both the Tevatron and the LHC, the experimental results agree well with the NNLO theoretical predictions.  The NNLO total cross section has been used to constrain the gluon PDF~\cite{Czakon:2013tha},
especially at large Bjorken scalling variable x, which plays a significant role in theoretical predictions of many NP scenarios. Besides, the  NNLO results can be used to improve NP studies, such as separating the stop signals from large top backgrounds in the stop searches~\cite{Czakon:2013xaa}.

\subsubsection{Threshold resummation}

When the physical process considered involves multi-scale in high energy hard scattering, in certain kinematic region, there exist the powers of large logarithms which origin from soft gluon effects so that the convergence of the fixed-order calculations in the QCD is spoiled. These large logarithms can be resummed by reorganizing the perturbative expansion, which is so-called soft gluon resummation.

The threshold resummation for top-pair production can be mainly divided into three different cases which are well reviewed in Ref.~\cite{Kidonakis:2011ca} and are summarized in Table~\ref{tab:soft_thresh}.  The soft gluon resummations for top-pair production at NLL accuracy have been available for a long time~\cite{Kidonakis:1997gm,Bonciani:1998vc}. The advances in the understanding of the infrared structure of QCD amplitudes~\cite{Becher:2009kw,Ferroglia:2009ep} make it possible to extend the resummation to NNLL level. In the threshold limit  $\beta\to 0$, the NNLL resummed total cross section was  calculated~\cite{Moch:2008qy}. The NNLL resummation for top quark pair invariant-mass~(PIM) distribution has been investigated~\cite{Ahrens:2010zv,Cacciari:2011hy}. Recently, NNLL resummation for the transverse-momentum and rapidity distributions of the top quark were also calculated in the case of single particle inclusive~(1PI) kinematics~\cite{Kidonakis:2010dk,Ahrens:2011mw,Ahrens:2011px}. Utilizing the
results of soft gluon resummation, the approximate NNLO
corrections are calculated in Refs.~\cite{Ahrens:2010zv,Ahrens:2011mw,Ahrens:2011px,Ferroglia:2012uy,Ferroglia:2013zwa,Ferroglia:2013awa,Kidonakis:2010dk,Cacciari:2011hy,Beneke:2009ye,Beneke:2010fm,Beneke:2011mq}.

Table~\ref{tab:results} shows the most accuracy predictions for the total cross section at NNLO+NNLL level~\cite{Czakon:2013goa}. Their scale uncertainties are about $2~\%$ and 4~\%  at the  Tevatron and the LHC, respectively. Compared with the NNLO results in Table~\ref{tab:NNLO},  the resummation results agree with the NNLO results and reduce the scale dependence.

\begin{table}[h]
\begin{center}
\caption{\small The NNLO+NNLL results for top-pair production at the Tevatron and the LHC~\cite{Czakon:2013goa}}
\begin{tabular}{c c c c}
\hline
Collider & $\sigma_{\rm tot}$ (pb)  & scales (pb) & pdf (pb) \\
\hline
Tevatron & $7.164$ & $ ^{+0.110 (1.5~\%)}_{-0.200 (2.8~\%)} $ & $ ^{+0.169 (2.4~\%)}_{-0.122 (1.7~\%)} $   \\
LHC 7 TeV & $172.0$ & $ ^{+4.4 (2.6~\%)}_{-5.8 (3.4~\%)} $ & $ ^{+4.7 (2.7~\%)}_{-4.8 (2.8~\%)} $  \\
LHC 8 TeV & $245.8$ & $ ^{+6.2 (2.5~\%)}_{-8.4 (3.4~\%)} $ & $ ^{+6.2 (2.5~\%)}_{-6.4 (2.6~\%)} $  \\
LHC 14 TeV & $953.6$ & $ ^{+22.7 (2.4~\%)}_{-33.9 (3.6~\%)} $ & $ ^{+16.2 (1.7~\%)}_{-17.8 (1.9~\%)} $   \\
\hline
\end{tabular}

\label{tab:results}
\end{center}
\end{table}

\subsubsection{Transverse momentum resummation}

\begin{figure}[h]
  \centering
  \includegraphics[scale=0.4]{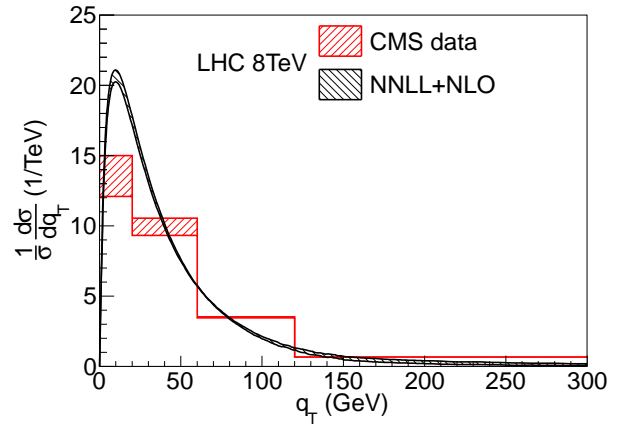}
  \vspace{-2ex}
  \caption{ \label{fig:res_exp} Comparison of normalized distribution between the resummation prediction and the data from the CMS at the 8 TeV LHC~\cite{Li:2013mia}}
\end{figure}

The transverse momentum distribution is one of the interesting observables for top-pair production. The measurements at the Tevatron show that the forward-backward asymmetry of $t\bar{t}$ production has strong dependence on the transverse momentum of $t\bar{t}$ system~\cite{Abazov:2011rq,Aaltonen:2012it}.
An enhancement of the sensitivity of the invariant mass distribution to the effects of NP can be obtained by setting a kinematic cut on the top quark pair transverse momentum, especially in the small $q_{\rm T}$ region~\cite{Alvarez:2012uh}. Therefore, it is significant to have an accuracy prediction for small $q_{\rm T}$ distribution in top-pair production.

It is well known that there are the large logarithms of the form $\ln^n (q_{\rm T}/M)$ at small $q_{\rm T}$ region in the fixed-order calculations. To obtain the correct prediction at small $q_{\rm T}$,  these logarithms must be resummed to all order in the QCD coupling constant $\alpha_{\rm s}$. Efforts have been made in order to achieve the transverse-momentum resummation by modifying the CSS formalism~\cite{Collins:1984kg}. However, they neglected the color-mixing effects between the singlet and octet final states and the contributions from the inial-final state soft-gluon exchange. Another approach to the transverse momentum resummation is based on the SCET, which has been developed for Drell Yan process and Higgs production~\cite{Gao:2005iu,Idilbi:2005er,Becher:2010tm,Becher:2011xn,GarciaEchevarria:2011rb,Chiu:2012ir,Becher:2012yn},
where there are not any colored particle in final state. Based on SCET, the NNLL transverse momentum resummation for top-pair production was obtained in Refs.~\cite{Zhu:2012ts,Li:2013mia}. Their factorization formula can be written as
\begin{widetext}
\begin{multline}
  \label{eq:master}
  \frac{{\rm d}^4\sigma}{{\rm d}q_{\rm T}^2 \, {\rm d}y \, {\rm d}M \, {\rm d}\cos\theta} = \sum_{i=q,\bar{q},g} \sum_{a,b} \frac{8\pi\beta_tM}{3s(M^2+q_{\rm T}^2)}
  \int_{\xi_1}^1 \frac{{\rm d}z_1}{z_1} \int_{\xi_2}^1  \frac{{\rm d}z_2}{z_2}  \, f_{a/N_1}(\xi_1/z_1,\mu) \, f_{b/N_2}(\xi_2/z_2,\mu)
  \\
  \times C_{i\bar{i} \leftarrow ab}(z_1,z_2,q_{\rm T},M,m_t,\cos\theta,\mu) \, ,
\end{multline}
with
\begin{multline}
  \label{eq:cii}
  C_{i\bar{i} \leftarrow ab} (z_1,z_2,q_{\rm T},M,\cos\theta,m_t,\mu) =
     \frac{1}{2} \int^\infty_0 x_{\rm T}{\rm d}x_T \, J_0(x_{\rm T}q_{\rm T})  \, \exp \big[ g_i(\eta_i,L_\perp,\alpha_{\rm s}) \big]
     \left[ \bar{I}_{i \leftarrow a}(z_1,L_\perp,\alpha_{\rm s}) \, \bar{I}_{\bar{i} \leftarrow b}(z_2,L_\perp,\alpha_{\rm s}) \right.
  \\ \left. +\delta_{gi} \bar{I}_{g \leftarrow a}^{\prime}(z_1,L_\perp,\alpha_{\rm s}) \, \bar{I}_{g\leftarrow b}^\prime(z_2,L_\perp,\alpha_{\rm s})\right] \,
    \mathrm{Tr} \Big[ \bm{H}_{i\bar{i}}(M,m_t,\cos\theta,\mu) \, \bm{S}_{i\bar{i}}(L_\perp,M,m_t,\cos\theta,\mu) \Big] .
\end{multline}
\end{widetext}
where $\bm{H}_{i \bar{i}}$ and $\bm{S}_{i\bar{i}}$  are the hard function and soft function, respectively. $J_0(x_{\rm T} q_{\rm T})$ is the 0th order Bessel function. The functions $g_i$, $\bar{I}_{\bar{i} \leftarrow b}$ and $\bar{I}^\prime_{\bar{i} \leftarrow b}$ are related to the transverse momentum dependent parton distribution functions~\cite{Becher:2010tm,Becher:2012yn}. The hard functions are matrices in the color space, which are the same as in threshold resummation~\cite{Ahrens:2010zv}. The soft functions are defined and calculated in Refs.~\cite{Zhu:2012ts,Li:2013mia}, which are given by
\begin{eqnarray}
  \bm{S}_{i\bar{i}}^{(1)} &&= 4 L_\perp \left( 2\bm{w}^{13}_{i\bar{i}}  \ln\frac{-t_1}{m_tM}  + 2\bm{w}^{23}_{i\bar{i}}
  \ln\frac{-u_1}{m_tM} + \bm{w}^{33}_{i\bar{i}} \right)
  \nonumber \\ &&
  - 4 \left( \bm{w}^{13}_{i\bar{i}} + \bm{w}^{23}_{i\bar{i}} \right) \mathrm{Li}_2 \Biggl( 1 - \frac{t_1u_1}{m_t^2M^2} \Biggr)
  + 4\bm{w}^{33}_{i\bar{i}} \ln\frac{t_1u_1}{m_t^2M^2}
    \nonumber  \\&&
  -  2\bm{w}^{34}_{i\bar{i}}
  \, \frac{1+\beta_t^2}{\beta_t} \, \bigl[ L_\perp \ln x_s + f_{34} \bigr] \, ,
\end{eqnarray}
with $x_s = (1-\beta_t)/(1+\beta_t)$ and
\begin{align}
  f_{34} =&  - \mathrm{Li}_2 \left( -x_s \tan^2\frac{\theta}{2} \right)
  + \mathrm{Li}_2 \left( -\frac{1}{x_s} \tan^2\frac{\theta}{2} \right)
      \nonumber \\  &
  + 4\ln x_s \ln\cos\frac{\theta}{2} \, .
\end{align}
In general, the renormalization group equations for the soft functions can be written as \begin{equation}
     \frac{\rm d}{{\rm d}\ln \mu} \bm{\mathcal{S}}_{i\bar{i}} = -\gamma_{i\bar{i}}^{s\dagger} \bm{\mathcal{S}}_{i\bar{i}}-
        \bm{\mathcal{S}}_{i\bar{i}}\gamma_{i\bar{i}}^s\ .
\end{equation}

Fig.~\ref{fig:res_exp} shows the transverse momentum distribution, which has been matched to QCD NLO results. It can be found that the resummed distributions have small scale dependence and are consistent with the data  from the CMS~\cite{CMS:fxa} within theoretical and experimental uncertainties.

%Expanding the resummed formula provides another approach to construct the subtraction terms for the QCD NNLO corrections to top quark pair production based on the $q_{\rm T}$ subtraction method~\cite{Catani:2007vq}.
This formalism can be used to calculate transverse momentum resummation for other massive colored particle production processes at hadron colliders.

\subsection{Single top quark production}

\begin{figure*}
   \centering
   \includegraphics[width=0.7\textwidth]{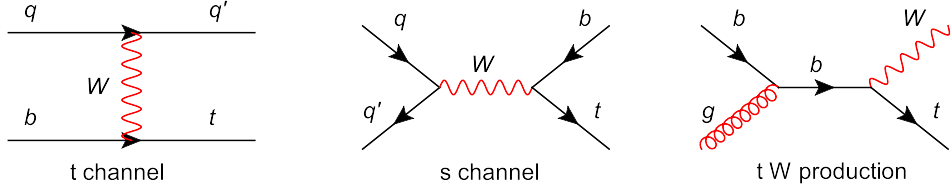}
   \vspace{-2ex}
   \caption{\label{fig:single_top}(Color online) Feynman diagrams for single top production}
\end{figure*}

\begin{figure*}
  \centering
  \includegraphics[width=0.38\textwidth]{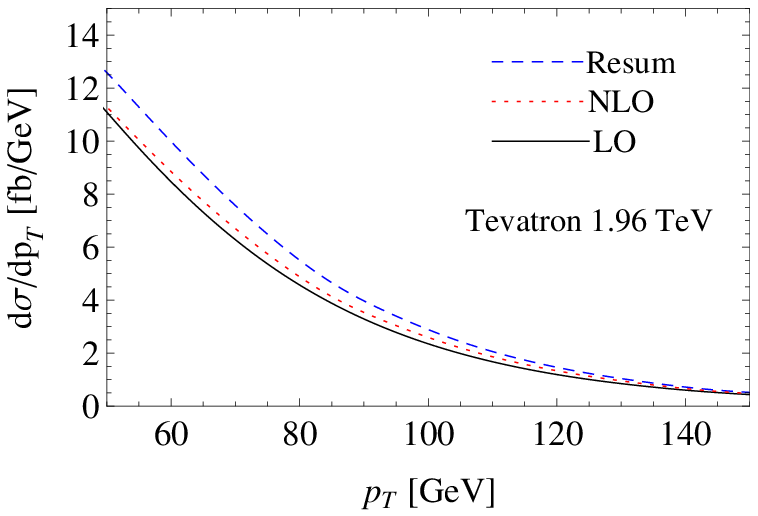} \ \ \
  \includegraphics[width=0.4\textwidth]{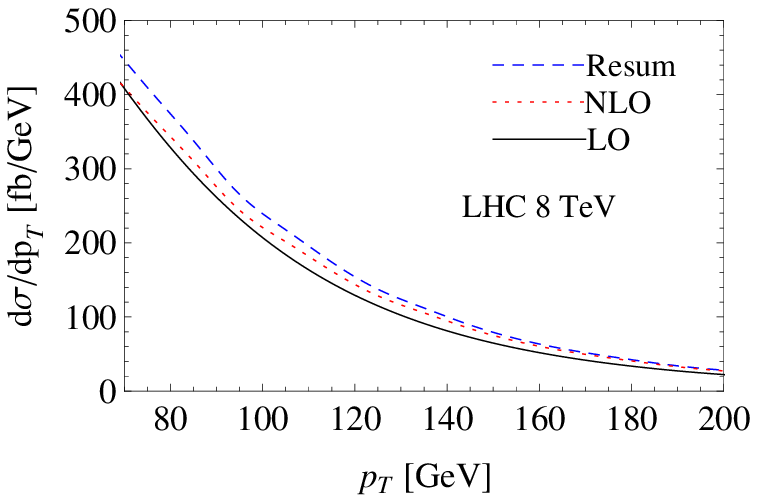}
  \vspace{-2ex}
  \caption{(Color online) The RG improved (dashed) and fixed-order $q_{\rm T}$ distributions for $t$-channel single top production at the Tevatron (left) and the LHC (right)~\cite{Wang:2012dc}}
  \label{fig:s-t-resum}
\end{figure*}

\begin{figure}[h]
 \centering
 \includegraphics[scale=0.4]{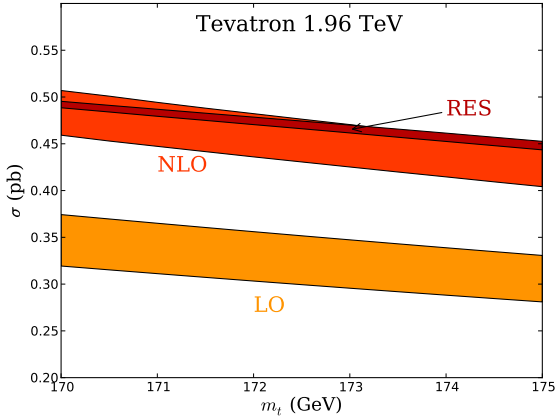}
\caption{\label{fig:s-top} (Color online) Top quark mass dependence of the fixed-order and
resummed cross section~\cite{Zhu:2010mr}}
\end{figure}

Besides top quark pair production, another interesting process is single top quark production, which was first observed at the Tevatron~\cite{Aaltonen:2009jj,Abazov:2009ii} and later at the LHC~\cite{Aad:2012ux,Chatrchyan:2012ep}.

The production of single top quark provides an unique window into the $V-A$ structure of the $Wtb$ vertex and a direct extraction of the CKM matrix element $V_{tb}$. Furthermore, the single top quark production is an important background in searching for NP. As shown in Fig.~\ref{fig:single_top}, in the SM, the single top production is divided into three channels: $t$-channel production, $s$-channel production and associated production of top quark and $W$ boson.

The QCD NLO corrections to single top production were investigated~\cite{Bordes:1994ki,Stelzer:1995mi,Smith:1996ij,Harris:2002md,Sullivan:2004ie} in the approximation of a stable top. The investigations of matching the QCD NLO corrections to parton shower were performed in Refs.~\cite{Frixione:2005vw,Alioli:2009je,Re:2010bp,Frederix:2012dh}. And single top production and decay at NLO was studied with the narrow width approximation in Refs.~\cite{Campbell:2004ch,Cao:2004ky,Cao:2004ap,Cao:2005pq,Heim:2009ku,Schwienhorst:2010je}. Off-shell effects for $t$-channel and $s$-channel single-top production were also calculated~\cite{Falgari:2010sf,Falgari:2011qa}. For $t$-channel single top production, the relation between the four favor scheme~(4F) and the five scheme~(5F) scheme  was studied~\cite{Campbell:2009ss,Campbell:2009gj}. The total cross sections in the 4F scheme is found to be smaller and with larger uncertainties than the ones in the 5F scheme. However, the predictions of the two schemes were found to be
in substantial agreement.
%They found that the cross section based on 4F scheme is consistent with that based on 5F scheme in  Ref..

Beyond the fixed-order calculations,  the soft gluon resummations improve the theoretical predictions. Among the three production channels at hadron colliders, the $t$-channel is the dominant one at both the Tevatron and the LHC. In the CSS framework, the NLL and NNLL threshold effects were calculated in Refs.~\cite{Kidonakis:2006bu,Kidonakis:2007ej,Kidonakis:2011wy}.  The top quark transverse momentum distribution at large $p_{\rm T}$ is interesting because it can be directly compared with the experimental results and is an important background in the searches of NP. This has been investigated with SCET in the partonic threshold limit $s_4 \to 0$~\cite{Wang:2012dc}, which is the first application of SCET to a spacelike process with the final states of one massless and one massive colored particle. In the SCET approach, the differential cross section at partonic level can be factorized into the convolution of hard, jet and soft functions, which can be written in the form~\cite{Wang:2012dc}
\begin{eqnarray}
      \label{eqs:facmain2}
       \frac{{\rm d}\hat{\sigma}_{ij}^{\rm thres}}{{\rm d}\hat{t}{\rm d}\hat{u}} &=&
             \frac{1}{4N^2_c} \frac{1}{8\pi}\frac{1}{\hat{s}}
             \lambda_{0,ij} H_{\rm up}(\mu) H_{\rm dn}(\mu)
             \int {\rm d}k^+\int {\rm d}p_1^2
                          \nonumber \\
             && \times \mathcal{S}(k^+,\mu) J(p_1^2,\mu)\delta(s_4-p_1^2-2k^+E_1),
\end{eqnarray}
where $H_{\rm up}$ and $H_{\rm dn}$ stand for contributions from the up and down fermion lines, respectively. The hard, jet and soft functions represent interactions at different scales, which can be calculated order by order in perturbative theory. After combining the hard, jet and soft functions, the RG improved top quark $p_{\rm T}$ distributions  are shown in Fig.~\ref{fig:s-t-resum}~\cite{Wang:2012dc}. It can be seen that the resummed distribution is increased by about $9~\% - 13~\%$ and $4~\% - 9~\% $ for $p_{\rm T} > 50$ and 70 GeV at the Tevatron and the LHC, respectively. Recently, the soft gluon resummation in the partonic threshold $s_4 \to 0$ was recalculated in the CSS framework, which also improved the NLO calculations by including soft-gluon corrections at NNLO~\cite{Kidonakis:2013yoa}.

As for the $s$-channel single top production, it is also an important process, because it is sensitive to the interaction mediated by an extra heavy particle.  Approximate NNLO calculations from the NLL and NNLL threshold resummation in the CSS framework were presented in Refs.~\cite{Kidonakis:2006bu,Kidonakis:2007ej} and Ref.~\cite{Kidonakis:2010tc}, respectively. Based on SCET, the factorization and the NNLL resummation results were given in Ref.~\cite{Zhu:2010mr}, where the cross section is also factorized into the convolution of hard, jet and soft functions:
\begin{eqnarray}
\label{fe}
  \sigma^{\rm thres} &&=
  \frac{1}{2E^2_{\rm CM}} \frac{1}{4N^2_{\rm c}}\int^1_0
  \frac{{\rm d}x_a}{x_a}\frac{{\rm d}x_b}{x_b} \int  \frac{{\rm d}^3q}{2 E_q
  (2\pi)^3}
  \nonumber \\ && \times
  f_{i/P_a}(x_a,\mu_f)
  f_{j/P_b}(x_b,\mu_f)
  \lambda_{0,ij} H_{IJ}
  \nonumber \\
  && \times
  \int {\rm d}k^+\, S_{JI}(k^+_i,\mu)(2\pi) J(s_4-2k^+
  E_1,\mu)\ .
\end{eqnarray}
Fig.~\ref{fig:s-top} shows the LO, NLO and resummed cross sections for different top quark mass at the Tevatron. Compared with the NLO results, it can be seen that  the scale dependence of the resummed cross section was significantly improved. Besides, the resummation effects enhance the NLO cross section by about $3~\% - 5~\%$.

The associated production of top quark with a  $W$ boson process $b g\to t W^-$ has the second largest cross section in the single top production at the LHC. The QCD NLO corrections were calculated  in Refs.~\cite{Giele:1995kr,Zhu:2001hw,Campbell:2005bb}. In the case of massive b-quarks, the NLO description of this channel (plus decay) was also investigated~\cite{Cascioli:2013wga}. Approximated NNLO  corrections from NLL and NNLL resummation were calculated in Refs.~\cite{Kidonakis:2006bu,Kidonakis:2007ej} and Ref.~\cite{Kidonakis:2010ux}, respectively. It is found that the approximate NNLO corrections increase the NLO cross by about $8~\%$.

\section{Summary}

We have briefly reviewed some recent theoretical processes on the high precision calculations in the Higgs boson and top quark physics at the hadron colliders, including the fixed-order and soft gluon resummation effects. The main aim of the future LHC experiments is precision test of the SM and search for the NP signal. Therefore, with the increasing of measurement accuracy at the LHC, it is a major task in future to exceed the present accuracy of the theoretical predictions and to perform higher order calculations for important processes,  in particular the processes involving Higgs boson and top quark, such as QCD N$^3$LO corrections to Higgs production, higher order QCD corrections to Higgs and jet associated production, the fully differential NNLO calculations for top-pair production, and the high order QCD calculations of top pair and jet associated production, which is a significant  background of SUSY signals, etc.

\begin{acknowledgments}
We would like to thank Jun Gao, Jian Wang and Hua Xing Zhu for useful suggestions. This work is supported by the National Natural Science Foundation of China under Grants No.~11375013 and No.~11135003.
\end{acknowledgments}

%\bibliographystyle{hunsrt}
%\bibliography{body/reference}

%\begin{thebibliography}{Tevatron Electroweak Working~Group13}

%\end{thebibliography}

\end{document}